\documentclass[
  journal=largetwo,
  manuscript=research-paper,
  year=2024,
  volume=37,
]{cup-journal}

\usepackage{amsmath}
\usepackage[nopatch]{microtype}
\usepackage{booktabs}
\usepackage{hyperref}
\usepackage{aas-macros}

\hypersetup{
    colorlinks=true,
    linkcolor=blue,
    filecolor=magenta,      
    urlcolor=cyan,
    pdftitle={Overleaf Example},
    pdfpagemode=FullScreen,
    }

\title[]{Enhanced Astrometry of the Rapid ASKAP Continuum Survey for Precise Localisation of Fast Radio Bursts}

\author{Akhil Jaini}
\affiliation{Center for Astrophysics and Supercomputing, Swinburne University of Technology, Post Office Box 218, Hawthorn, VIC 3122, Australia}
\email[Akhil Jaini]{ajaini@swin.edu.au}

\author{Adam T. Deller}
\affiliation{Center for Astrophysics and Supercomputing, Swinburne University of Technology, Post Office Box 218, Hawthorn, VIC 3122, Australia}
\alsoaffiliation{ARC Centre of Excellence for Gravitational Wave Discovery (OzGrav), Post Office Box 218, Hawthorn, VIC 3122, Australia}

\author{Yuanming Wang}
\affiliation{Center for Astrophysics and Supercomputing, Swinburne University of Technology, Post Office Box 218, Hawthorn, VIC 3122, Australia}
\alsoaffiliation{ARC Centre of Excellence for Gravitational Wave Discovery (OzGrav), Post Office Box 218, Hawthorn, VIC 3122, Australia}

\author{Emil Lenc}
\affiliation{CSIRO Space and Astronomy, Post Office Box 76, Epping, NSW 1710, Australia}

\author{Marcin Glowacki}
\affiliation{International Centre for Radio Astronomy Research (ICRAR), Curtin University, Bentley, WA 6102, Australia}
\alsoaffiliation{Institute for Astronomy, University of Edinburgh, Royal Observatory, Edinburgh, EH9 3HJ, United Kingdom}


\begin{document}

\begin{abstract}
Fast Radio Bursts (FRBs) are short, intense radio signals from distant astrophysical sources, and their accurate localisation is crucial for probing their origins and utilising them as cosmological tools. This study focuses on improving the astrometric precision of FRBs discovered by the Australian Square Kilometre Array Pathfinder (ASKAP) by correcting systematic positional errors in the Rapid ASKAP Continuum Survey (RACS), which is used as a primary reference for ASKAP FRB localisation. We present a detailed methodology for refining astrometry in two RACS epochs (RACS-Low1 and RACS-Low3) through crossmatching with the Wide-field Infrared Survey Explorer (WISE) catalogue.  The uncorrected RACS-Low1 and RACS-Low3 catalogues had significant astrometric offsets, with all-sky median values of $0.58''$ in RA and $-0.26''$ in Dec. (RACS-Low1) and $0.29''$ in RA and $1.24''$ in Dec. (RACS-Low3), with a substantial and direction-dependent scatter around these values. After correction, the median offset was completely eliminated, and the 68\% confidence interval in the all-sky residuals was reduced to $0.2''$ or better for both surveys. By validating the corrected catalogues against other, independent radio surveys, we conclude that the individual corrected RACS source positions are accurate to a 1-$\sigma$ confidence level of $0.3''$ over the bulk of the survey area, degrading slightly to $0.4''$ near the Galactic plane. This work lays the groundwork to extend our corrections to the full RACS catalogue that will enhance future radio observations, particularly for FRB studies.
\end{abstract}


\section{Introduction} \label{sec:introduction}

Fast Radio Bursts (FRBs) are millisecond duration single pulses of intense coherent radiation. These elusive bursts are discovered serendipitously, and offer a unique window into the universe. In recent years, the localisation of the first FRBs to galaxies far beyond our own has helped to establish their extreme luminosity and provided crucial insights into their potential origins (\citealp{2017Natur.541...58C}, \citealp{2019Sci...365..565B}). The detection of such powerful emissions from vast distances poses significant challenges to our understanding of the processes that generate radio waves and the kinds of astrophysical objects capable of producing such short-lived pulses. The most plausible explanation for these bursts is that they originate from neutron stars (\citealp{2019PhR...821....1P}), with theoretical and observational evidence pointing towards both magnetars (\citealp{2020ApJ...897....1L}, \citealp{2020Natur.587...59B}, \citealp{2020Natur.582..351C}) and high-mass X-ray binaries (\citealp{2020ApJ...893L..39L}) as potential sources. This suggests that multiple objects or mechanisms might be responsible for producing these enigmatic bursts. Studying FRBs is also significant because they can uniquely probe the universe's structure by detecting and characterising the highly diffuse intergalactic and circumgalactic media. These bursts traverse entire columns of plasma, revealing details about their densities, turbulence, and magnetisation. Measurements like dispersion measure (DM) and de-dispersed pulse width also provide valuable insights into the nearly invisible gas that makes up $\sim$ 90\% of baryonic matter in the universe (\citealp{2020Natur.581..391M}). 




\subsection{FRB Localisation} \label{sec:frblocalisation}
Since the discovery of the first FRB, astronomers have identified over 2000 unique sources of FRBs (\citealp{2022A&ARv..30....2P}), yet the precise localisation of these events remains a challenge, with just over 60 successfully linked to a host galaxy to date. Accurate astrometry is paramount in unravelling the enigmatic origins of FRBs and leveraging them as potent astrophysical and cosmological probes. For local universe galaxies with redshifts of $\sim 0.1$, an astrometric precision of a few arcseconds is sufficient to robustly identify a host galaxy, but for farther galaxies at redshifts $\sim 1$, we need sub-arcsecond astrometry for pinpointing the exact galactic and stellar environments housing FRB sources. 


Precise localisations are crucial for identifying FRB host galaxies, studying their properties and environments, and distinguishing between competing progenitor models. They also enable detailed mapping of the intergalactic medium (IGM) and its magnetic fields, enhancing our understanding of cosmic magnetism and large-scale structure. Furthermore, such localisation facilitates testing fundamental physics and cosmological principles, offering insights into dark matter, dark energy, and the early universe. Combining multi-wavelength observations with statistical analyses of localised FRBs provides valuable constraints on their physical properties and cosmological significance (\citealp{2022Sci...378.3043B}).

The fleeting nature of FRBs generally limits the options available in terms of precision calibration of the observations in which they are discovered. Particularly for non-repeating bursts, accurate localisation relies heavily on measuring their positions relative to background radio sources captured in the same interferometric observation. Consequently, by improving the precision of the astrometry for these continuum sources, which are generally catalogued in large-scale telescope surveys, we can enhance the accuracy of FRB localisation.

\begin{table*}[h]
    \centering
    \caption{RACS Observing Details}
    \begin{tabular}{|c|c|c|c|c|c|c|}
        \hline
        \textbf{Epoch} & \textbf{Year} & \textbf{Central} & \textbf{Bandwidth (MHz)} & \textbf{Number of}  & \textbf{Footprint} & \textbf{Reference} \\
         & & \textbf{Frequency (MHz)} & & \textbf{Scans} & & \\
        \hline
        RACS-Low1 & 2019-2020 & 887.5 & 288 & 903 & $square\_6x6$ & \citealp{2020PASA...37...48M} \\
        RACS-Low3 & 2023-2024 & 943.5 & 288 & 1493 & $closepack36$ & -\footnote{RACS-Low3 does not have a dedicated publication but follows similar methodologies to those outlined in \citealp{2021PASA...38...58H} and \citealp{2024PASA...41....3D}. Its documentation is available on the RACS website at: \url{research.csiro.au/racs/home/survey/comparison/} and \url{research.csiro.au/racs/racs-low3-dr1-raw-data/}.} \\
        RACS-Mid1 & 2020-2021 & 1367.5 & 144 & 1493 & $closepack36$ & \citealp{2023PASA...40...34D} \\
        RACS-High1 & 2022-2023 & 1655.5 & $\sim$ 200 & 1493 & $closepack36$ & \citealp{2024PASA...41....3D} \\
        \hline
    \end{tabular}
    \label{tab:racsepochs}
\end{table*}

\subsection{ASKAP} \label{sec:askapfrbs}
The Australian Square Kilometre Array Pathfinder (ASKAP) is an interferometer which comprises of 36, 12~m diameter antennas with a maximum baseline length of 6~km. Each antenna is equipped with novel Phased-Array Feed (PAF) receivers that, together with advanced digital systems, generate 36 simultaneous, independent beams, which collectively cover a $\sim30$~sq.deg. field-of-view in the frequency range of 700-1800~MHz (\citealp{2021PASA...38....9H}). These beams are calibrated independently and can produce separate (albeit overlapping) images, but are commonly mosaicked together to form a single large image.

ASKAP produces correlated visibilities with an integration time of 10~s and frequency resolution of 18~kHz. This data product is used for rapid surveying of the sky in continuum and spectral line processing modes. ASKAP's wide field-of-view has helped discover various transient phenomena, including flare stars, long period transients (LPTs), and, most notably, FRBs. However, the utility of ASKAP extends beyond mere detection. The interferometric capabilities of ASKAP allow for sub-arcsecond astrometry, which is crucial not only for precisely localising FRBs but also for other transient phenomena. For flare stars and LPTs, precise localisation is critical for cross-matching against crowded optical catalogues, particularly when only a single detection has been made. This ensures that the sources can be accurately identified and followed up across other wavelengths, allowing astronomers to explore their physical properties and origins comprehensively. Therefore, the results of this work, which enhance the localisation precision of ASKAP detections, are highly relevant to a broader range of data products beyond just FRBs.


\subsection{FRB detection and localisation with ASKAP}

ASKAP has two FRB detection pipelines which operate commensally with the standard processing. The first one uses the InCoherent Sum (ICS, started in mid-2018, \citealp{2024arXiv240802083S}) of antenna powers at a time resolution of $\approx 1$~ms and frequency resolution of 1~MHz. Each beam is independently summed and dedispersed. The sensitivity of the ICS processing is proportional to $N_{\rm ant}^{1/2}$, where $N_{\rm ant}$ is the number of antennas.

The second pipeline, Commensal Real-time ASKAP Fast Transients (CRAFT) COherent (CRACO, \citealp{2024arXiv240910316W}) Upgrade, commenced operations in late 2023. CRACO can potentially utilise interferometric visibilities with a time resolution of $\approx 1$~ms and frequency resolution of $1$~MHz. It creates and dedisperses an interferometric image coherently, with sensitivity scaling proportionally to $N_{\rm ant}$. Currently, CRACO operates with a time resolution of 13.8 ms, meaning its sensitivity is only comparable to ICS for FRBs that are unresolved in time. However, once CRACO achieves its targeted 1~ms time resolution, the detection sensitivity is expected to improve by a factor of $\sim 5$ over ICS, significantly enhancing its capability for detecting FRBs.


Both detection systems currently operate in real time. When an FRB is detected, the system saves the voltage data to disk, which is then correlated offline, and interferometric calibration and imaging, as well as high-time resolution processing is performed as described in \citealp{2023A&C....4400724S}. In addition to imaging the $\sim$milliseconds of time containing the FRB itself, the entire download (3--12 seconds) is imaged, producing a radio continuum image of the entire field with an RMS sensitivity of a few milli-Jansky. We measure an astrometric offset by comparing the positions of the sources in this image with a reference catalogue. After measuring the position of the FRB from the $\sim$millisecond-duration image containing it, we adjust FRB position by the astrometric offset derived from the field sources. 






\subsection{RACS} \label{sec:askapracs}
The Rapid ASKAP Continuum Survey (RACS) is a comprehensive all-sky survey conducted using ASKAP. It represents ASKAP's pilot survey and aims to create a shallow model of the ASKAP sky to aid in the calibration of future deep ASKAP surveys and other scientific research. RACS covers the entire southern sky below the Dec. of $+49^\circ$ across the full ASKAP band of 700–1800 MHz (\citealp{2020PASA...37...48M}). This survey aims to provide a comprehensive model of the radio sky accessible to ASKAP to modest depth, but significantly better than previous all-sky surveys in the southern hemisphere. The RACS project encompasses the entire sky visible from the ASKAP site in Western Australia, and the initial data release includes 903 images providing information for over 3 million continuum sources, and serves as a critical reference for subsequent astronomical investigations.



RACS source lists are created for each observing run, known as an SBID (Scheduling Block Identifier). For each SBID, a mosaic image (also known as a tile image) is produced by combining data from 36 individual ASKAP beams, which overlap to cover a large area of the sky. From these tile images, source lists of detected radio sources are extracted, with key parameters like flux density, position, and source morphology. These source lists are then compiled into catalogues after undergoing further checks (as described in \citealp{2021PASA...38...58H}).


Each RACS field (also referred to as scan) has a field-of-view of approximately 31~sq. deg. Each observation within RACS is made for 15~minutes and employs a contiguous 288 MHz band (unless mentioned otherwise) within the broader 700-1800 MHz range, with the first data release centred at 887.5 MHz. The survey achieves an angular resolution of approximately $15''$ and aims for a point-source detection sensitivity of around 1 mJy. RACS offers significant advantages over existing radio surveys like the National Radio Astronomy Observatory's Very Large Array (NRAO VLA) Sky Survey (NVSS, \citealp{1998AJ....115.1693C}) and the Sydney University Molonglo Sky Survey (SUMSS, \citealp{1999ldss.work..103S}), including greater depth, superior spatial resolution, sensitive wide-band coverage in the intermediate frequency regime for studying broadband spectra and transient behaviour, and publicly accessible data products such as total-intensity images, polarisation data, and a comprehensive source catalogue. RACS is also the only survey to uniformly cover the entire Southern sky at high angular resolution. Due to these advantages and overlapping sky coverage with other ASKAP data products, the positions of FRBs discovered with ASKAP are corrected using the continuum source positions provided by the RACS survey.

\subsubsection{RACS Bands} \label{sec:racsepochs}
The ASKAP observing band is split into three bands and RACS covers each of these bands (abbreviated as RACS-Low, RACS-Mid and RACS-High) in dedicated epochs, as detailed in Table~\ref{tab:racsepochs}. 




\subsubsection{Astrometry in RACS} \label{sec:racsastrometry}
RACS employs an observing setup where multiple target scans share a single common scan on a bright calibrator to provide initial phase, gain, and bandpass solutions (\citealp{2020PASA...37...48M}). However, any residual time- or direction-dependent errors between the calibrator and target scans can introduce astrometric errors into the initial RACS field model since no further phase calibration was applied prior to imaging and self-calibration. These errors can thus, propagate through to the final RACS catalogue positions. When compared to the International Celestial Reference Frame (ICRF), RACS sources exhibited notable systematic position offsets of magnitudes $\sim0.6''$ in right ascension and $\sim0.4''$ in declination ({\citealp{2020PASA...37...48M}}), and have an astrometric precision of less than $0.8''$. 

In this paper, we address and correct these astrometric uncertainties, ensuring that ASKAP-discovered FRBs can be localised with high precision, thus facilitating more accurate scientific investigations.




\subsection{Astrometry in Reference Catalogues} \label{sec:astrometry}
To test and correct the astrometry of RACS, we need to compare it with another catalogue. Ideally, this catalogue would have the same spatial resolution as RACS, be conducted at the same radio frequency, cover the entire RACS footprint, and have negligible astrometric errors. However, no such catalogue exists. Instead, we make use of several wide-field catalogues, each with its own advantages and disadvantages. We consider four potentially useful catalogues for both the astrometric correction and validation of RACS. These are the VLA FIRST (the Very Large Array Faint Images of the Radio Sky at Twenty-cm,\citealp{1994ASPC...61..165B}), VLASS (the Very Large Array Sky Survey, \citealp{2020PASP..132c5001L}), RFC (the Radio Fundamental Catalogue\footnote{Website: \url{astrogeo.org/rfc/}}, \citealp{2024arXiv241011794P}), and NASA WISE (the National Aeronautics and Space Administration's Wide-field Infrared Survey Explorer, \citealp{2010AJ....140.1868W}). The important features of these catalogues are detailed in Table~\ref{tab:refcatalogues}.

\begin{table*}[h!]
    \centering
    \caption{Survey Characteristics of VLA FIRST, VLASS, RFC and NASA WISE}
    \begin{tabular}{|c|c|c|c|c|c|}
        \hline
        \textbf{Survey} & \textbf{Wavelength/} & \textbf{Angular} & \textbf{Sky Coverage} & \textbf{Number of} & \textbf{Key Features} \\
         & \textbf{Frequency} & \textbf{Resolution} &  & \textbf{Sources} & \\
        \hline
        VLA FIRST & 1.4 GHz & 5 arcsec & $\sim$10,000 sq. deg. & $\sim$900,000 & Radio survey focusing on extragalactic \\
         &  &  & (Galactic caps) & & objects like quasars, radio galaxies \\
        \hline
        VLASS & 2-4 GHz & 2.5 arcsec & 33,885 sq. deg. (North of & $\sim$5,000,000 & Designed for transient radio phenomena \\
         &  &  & declination $-$40 deg.) &  & such as supernovae, black hole jets, etc. \\
        \hline
        RFC & Various (VLBI) & Milliarcsec & Entire sky & $\sim$22,000 & Provides astrometric data, used for \\
         &  &  &  &  & differential astrometry, phase calibrations \\
        \hline
        NASA WISE & 3.4, 4.6, & 6.1 - & Entire sky & $\sim$563,000,000 & Infrared sky survey covering asteroids, \\
         &  12, 22 microns & 12 arcsec &  &  & star formation, galaxy structure \\
        \hline
    \end{tabular}
    \label{tab:refcatalogues}
\end{table*}

The sources catalogued by these surveys are predominantly galaxies, which provide well-defined radio and/or infrared emissions, allowing accurate positional cross-matching. For any individual galaxy being cross-matched, the goal is to identify a corresponding source in the reference catalogues and compare the positions. Astrometry in these catalogues is, therefore, critical for accurately determining the positions of celestial objects. The FIRST survey, conducted with the VLA, has systematic astrometric errors of less than $0.05''$ and the source positions are measured to a 1-$\sigma$ statistical precision of $0.5''$ at the survey threshold of $1$~mJy, and much smaller for brighter sources. However, it covers only a small portion of the sky. 

VLASS is a more recent survey using the VLA, at a higher frequency and a higher angular resolution. Nominally, this should lead to a statistical astrometric precision of $0.25''$ for brighter sources, but due to imperfect widefield imaging corrections, the current Quick Look (QL) images have an astrometric precision ranging from $0.5''$ to $1''$. However, it only covers a part of the RACS sky, and the higher frequency of observation means the sources are more dominated by AGN cores whereas RACS-Low tends to pick up more radio lobes. The published catalogues of VLASS are also known to have a systematic astrometric error of $\approx 0.25''$ in declination (\citealp{2021ApJS..255...30G}). 

RFC is a database of compact radio sources offering milliarcsecond (mas) precision in source positions, achieved through Very Long Baseline Interferometry (VLBI). VLBI combines data from widely separated radio telescopes to form a virtual telescope with an effective aperture equal to the distance between them, enabling very high angular resolution. While RFC provides highly accurate astrometry, it only contains data on $\sim22,000$ sources (\citealp{2024arXiv241011794P}). Additionally, some of these sources exhibit extended structure on larger scales ($\sim$ arcseconds) that are resolved out by the VLBI observations. This means that the milliarcsecond positions provided by VLBI do not always represent the centroid of the larger-scale emission and may not align perfectly with the ASKAP positions. 

Finally, being a space-based infrared survey, WISE is not subject to the atmospheric and ionospheric propagation effects that perturb instrumental effects as the other radio interferometric surveys in this study, and the astrometric accuracy is typically better than $0.2''$, but it is observed in a completely different band (which means WISE primarily picks up AGN cores and not radio lobes like RACS-Low). 



Given these constraints, we chose to use WISE as the primary reference catalogue and while different frequency bands can cause centroid position offsets between WISE and RACS-Low for individual sources, since the Universe lacks a preferred direction, these positional offsets are statistically expected to average out across a large sample of sources within a RACS beam (as discussed in Section~\ref{sec:modelracslow1}). 


\bigskip

Section~\ref{sec:methodology} describes the techniques employed in this project, including the modelling parameters and their explanations, the governing equations and the main assumptions. Section~\ref{sec:results} presents the quantitative results, with separate subsections for results 
related to different objectives and different special cases. In Sections~\ref{sec:discussion} and ~\ref{sec:conclusion}, we discuss the future scope of the project, highlighting the practical possibilities and related considerations, and end with concluding remarks.

\section{Methodology} \label{sec:methodology}

This section elucidates the techniques employed in this project. The entire modelling pipeline\footnote{GitHub repository: \url{github.com/jainiakhil/RACS_Astrometry}} was developed in Python version 3.11, incorporating relevant modules such as \textit{astroquery} (\citealp{2019AJ....157...98G}), \textit{astropy, scipy}, and \textit{numpy} (\citealp{harris2020array}). 


\begin{figure}[h!]
    \centering
    \includegraphics[scale=0.61]{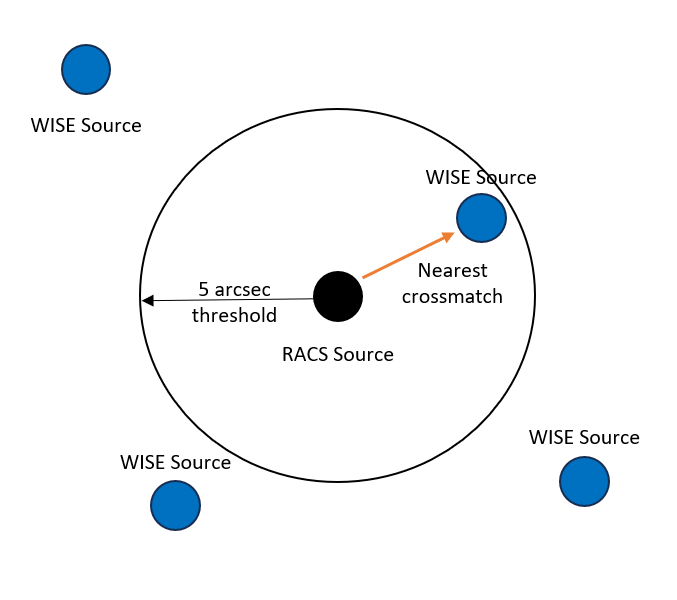}
    \caption{Graphically showing the crossmatching of a RACS source to a reference source (WISE, in this case) that is within $5''$ of its position. If there are more than 1 reference sources within the $5''$ threshold, the corresponding RACS source is not taken into consideration to avoid ambiguity in the calculations.}
    \label{fig:crossmatching}
\end{figure}

\begin{figure*}[h!]
    \centering
    \includegraphics[scale=0.62]{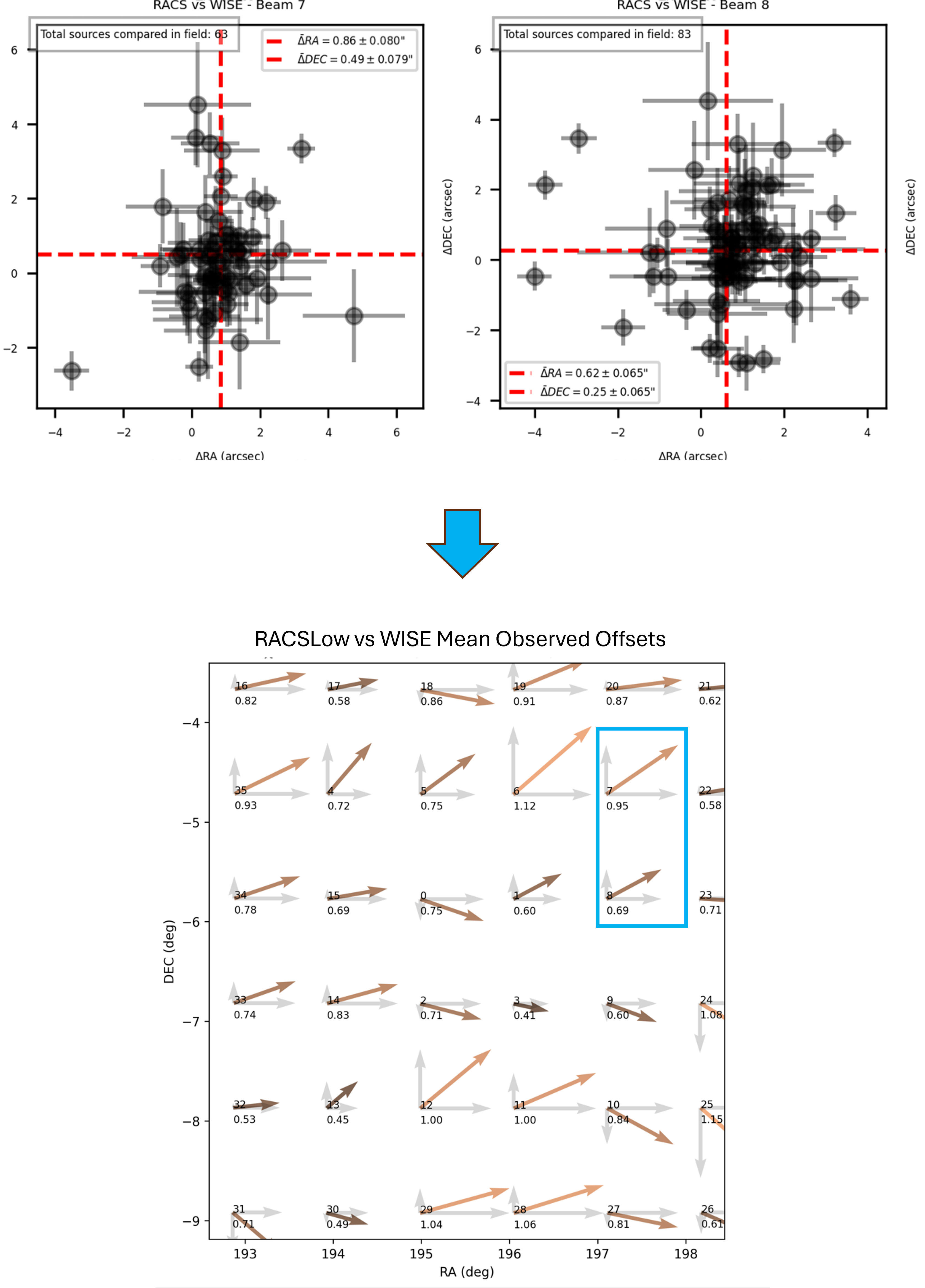}
    \caption{The mean observed offsets for a single scan are shown here. The top image shows the crossmatched offsets for each source for beams 7 and 8 of the 36 beams which are then used to calculate the mean offsets and uncertainties, with the x- and y-axes representing RA and Dec. offsets in arcsec respectively; while the bottom image shows the quiver plot showing the mean offset magnitudes and directions for all 36 beams, with the corresponding beams 7 and 8 highlighted.}
    \label{fig:meanoffsets}
\end{figure*}

\subsection{Modelling Parameters for RACS-Low1} \label{sec:modelracslow1}

The aim of this work is to derive accurate astrometric corrections for every source in RACS-Low. Directly correcting the errors in each individual source position based on their offsets between WISE and RACS is not feasible: intrinsic differences in their centroid positions would result in new sources of error being added that would likely outweigh the corrections being made. Instead, we developed a model to estimate the underlying astrometric corrections by averaging across as many sources as was reasonable, ensuring that we minimise introducing any additional noise.


To generate the initial offsets for RACS-Low1, we conducted a comparison of the sources within each beam of the available RACS-Low1 scans against the WISE catalogue. Out of the total 903 scans, only 805 were publicly available, with those located above Dec. $+30^\circ$ being excluded. These scans were available as a sources-only catalogue and a Gaussian components catalogue, where the latter categorised individual Gaussian-shaped components of extended sources, which is crucial for detailed studies of complex radio sources that are not well-represented by a single point source. We chose to use the Gaussian components catalogue for our work.


The different observing frequencies mean that the centroid positions in infrared (WISE) and radio (RACS-Low) may differ for any individual source. However, since the Universe lacks a preferred direction, the positional offsets caused by differences in centroid placement across the electromagnetic spectrum are expected to average incoherently over the large sample of sources within a RACS beam, thus minimally impacting the overall astrometric correction (e.g., for a field with 100 crossmatched sources each having a typical offset of $1''$, the mean residual error will be reduced to the $0.1''$ level). Nevertheless, since centroid differences could be expected to be larger on average for significantly extended sources, we implemented a filtering algorithm designed to exclude obvious extended sources. This algorithm identifies point-like sources by ensuring that the ratio of integrated flux to peak flux remains below a predetermined threshold (of 1.5, in this case), and ensuring that the sources do not have multiple Gaussian components within $30''$ of each other. This would filter out RACS sources that have extended radio lobe components or that are compact doubles, where the WISE source would not be associated with either of the RACS components. Additionally, we filtered out sources with a signal-to-noise ratio (SNR) below 6-$\sigma$. Collectively, these steps ensure that only isolated, point-like sources, having a minimum SNR of 6-$\sigma$, were included in the astrometric corrections. Through these measures, we effectively minimised the impact of observing frequency-based centroid discrepancies on the mean positions of sources.




In order to estimate the systematic position error in the catalogue at a given position (generally, centred on the location of an individual ASKAP beam during a given RACS scan), we must select an ensemble of sources to compare against the reference catalogue. We select all sources within a specified radius of the beam centre, where the choice of radius balances statistical precision (a larger radius yields more sources and hence higher precision) against fidelity (a smaller radius minimises contamination of the offset by the mosaicking process that combines data from adjacent beams - see Section~\ref{sec:modelRACS-Low3} for more details). We tested radii between $0.5^\circ$ and $2^\circ$, finding that a $0.5^\circ$ radius yielded too few crossmatches, while a $2^\circ$ radius introduced excessive overlap with neighbouring beams. Ultimately, a $1^\circ$ radius provided the best compromise, offering sufficient crossmatches without significant overlap, and was adopted as the standard.

For each $1^\circ$ beam, we applied a crossmatching threshold of $5''$, chosen to minimise false positives while accommodating typical astrometric uncertainties. With an average WISE source density of approximately 13,500 sources per square degree, this threshold was expected to perform well, as the average separation between sources is about $31''$. While this strategy worked effectively for off-plane regions, the higher source density in the Galactic plane, combined with the prevalence of extended sources, resulted in increased false positives and decreased reliability.

The crossmatches were performed using \textit{astropy's} built-in \textit{match\_to\_catalog\_sky} function (\citealp{astropy:2013}, \citealp{astropy:2018}, \citealp{astropy:2022}). If multiple WISE sources fell within the crossmatching threshold of a single RACS source, we excluded that RACS source to avoid ambiguity in source identification (as shown graphically in Figure~\ref{fig:crossmatching}). This ensured that each crossmatch was a unique pairing, preventing errors in astrometric corrections. Given the high source density in some regions, this step was crucial to maintain precision, and since many sources survived this cut, discarding a few at this stage did not impact the overall analysis.


Quantitatively, each 1-degree radius RACS beam contains around 200 sources. The filtering algorithm described above eliminates roughly 30–50\% of these, reducing the count to about 120 sources eligible for crossmatching. Once crossmatched, an additional 30–40\% of sources are typically removed, resulting in around 80 matched sources per beam. However, in the Galactic plane regions, the filtering process removes a larger fraction of sources due to increased density and complexity, resulting in fewer crossmatched sources.


This procedure was systematically applied to every RACS source within the entire sky coverage, for each beam, and for each scan, resulting in a comprehensive list of mean offsets and associated uncertainties for RA and Dec. positions.

Figure \ref{fig:meanoffsets} illustrates the beam-wise offsets observed in a specific RACS scan compared to WISE. These offsets range from a few milliarcseconds to several arcseconds, consistent with the results presented in \citealp{2020PASA...37...48M}, thereby exacerbating the localisation uncertainty of all past ASKAP FRBs. 


\begin{figure*}[h!]
    \centering
    \includegraphics[scale=0.9]{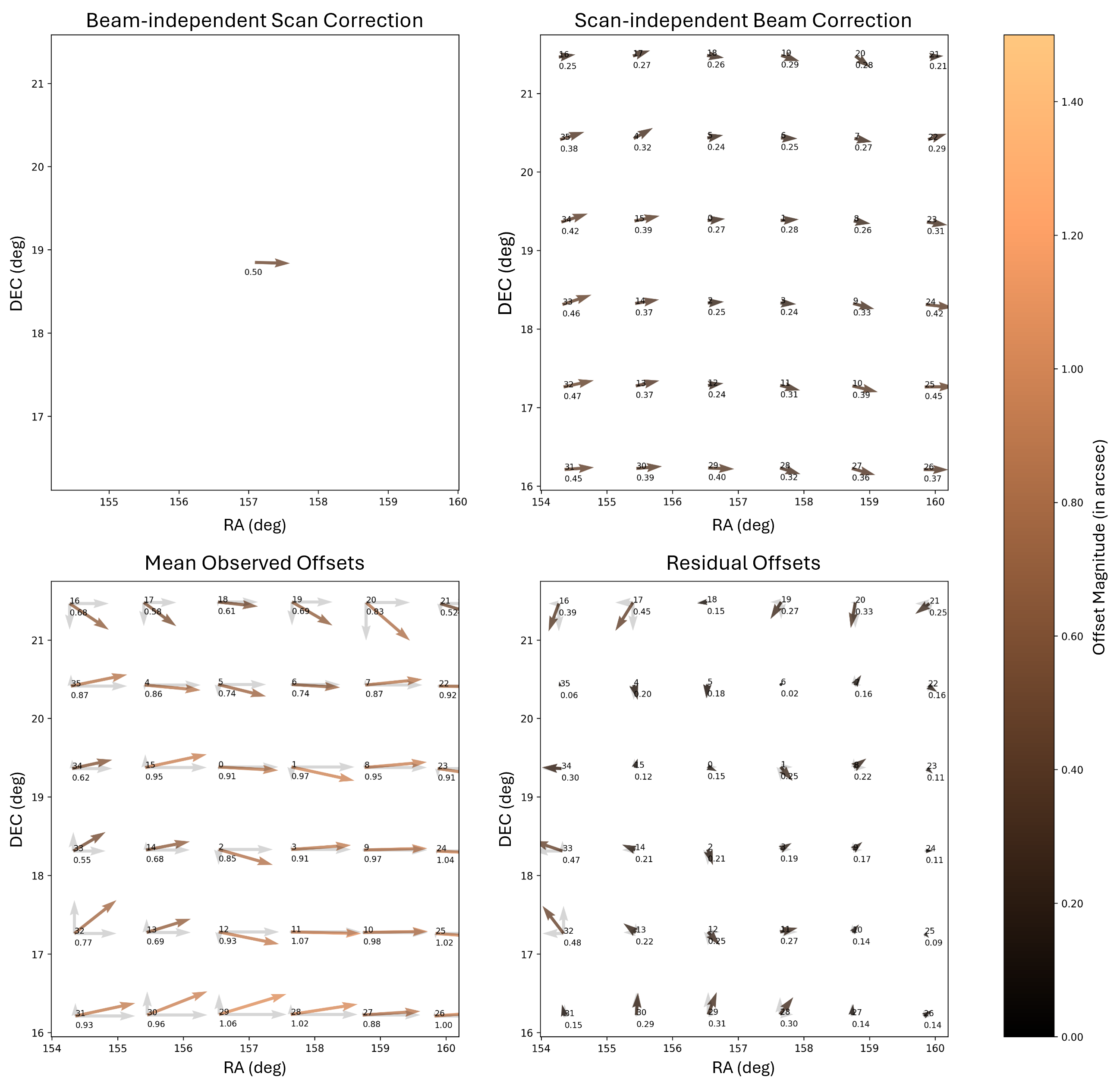}
    \caption{The observed, modelled and residual beam offsets for a particular RACS-Low1 scan (1028+18A) are shown here. The top-left plot shows the beam-independent scan offset modelled for the particular scan, the top-right plot shows the scan-independent beam offset modelled for all scans having the same bandpass calibration. The bottom-left plot shows the observed mean offsets for the scan, and the bottom-right plot shows the residual offsets after subtracting the sum of the beam-independent scan offset and the scan-independent beam offset from the observed mean offsets. As can be inferred visually and by referring to the offset magnitudes printed on the plots, the residual offsets are significantly lower than the observed offsets.}
    \label{fig:modelling}
\end{figure*}


In order to model the mean offsets as accurately as possible, we had to break the RACS dataset into smaller subsets having a common offset. We knew that the calibration process for RACS can introduce time- and direction-dependent calibration errors (as described in Section 2.7 of \citealp{2020PASA...37...48M}). Notably, the bandpass calibration (which is undertaken once per beam on a very bright calibrator) can lead to a common error across all scans using the same calibration, referred to as the {\em scan-independent beam offset} ($\Delta^i$), which represents the astrometric error in the direction of the calibrator at the time of the calibrator scan. Additionally, calibration terms that vary with direction and/or time (since subsequent scans can be many hours and many tens of degrees away from the bandpass calibrator) can lead to errors that are approximately common to all beams within a given scan, referred to as the {\em beam-independent scan offset} ($\Delta^j$). We note that $\Delta$ is used throughout this manuscript to refer to an offset, but this can be for an individual source, the average of an ensemble of sources, or as a fitted model parameter -- sub- or super-scripts are used to differentiate between these cases.

The total astrometric correction for any beam in a given scan can be defined: 


\begin{equation}    \label{eq:modeloffset}
    \Delta_{modelled}^{ij} = \Delta_{modelled}^{i} + \Delta_{modelled}^{j}
\end{equation}

\begin{equation}    \label{eq:residualoffset}
    \Delta_{residual}^{ij} = \Delta_{observed}^{ij} - \Delta_{modelled}^{ij}
\end{equation}

where, $\Delta_{observed}^{ij}$ is the observed offset for the $i$-th scan and $j$-th beam, $\Delta_{modelled}^{i}$ is the {\em beam-independent scan offset} and $\Delta_{modelled}^{j}$ is the {\em scan-independent beam offset}, and $\Delta_{residual}^{ij}$ is the residual offset after corrections, in both RA and Dec. The offset modelling was performed using the built-in \textit{scipy} function \textit{optimize.minimize} (\citealp{2020SciPy-NMeth}). Figure \ref{fig:modelling} provides an example of the offset modelling for a specific scan. This modelling process was subsequently applied to all RACS-Low1 scan sets.


Next, we introduced an error floor when calculating the uncertainties in RA and Dec. The uncertainty in the estimated offset for each coordinate was determined by taking a weighted sum of the observed uncertainties of both the target source and the reference source. For very bright sources, the RACS and WISE positions can be measured with very low uncertainties; however, the offset between their centroids might not be zero even for such sources. This non-zero intrinsic offset can arise due to differences in frequency coverage, resolution, and any subtle morphological mismatches between the RACS and WISE catalogues. In contrast, the {\em RACS position error} leads to systematic offsets that would average out over an ensemble of sources when comparing their RACS and WISE positions.

To account for this intrinsic offset and avoid underestimating the positional uncertainties for bright sources, we added a floor value in quadrature to the statistical offset uncertainty: 

\begin{equation}    \label{eq:flooradd}
    \sigma_\Delta = \sqrt{\sigma_t^2 + \sigma_r^2 + f^2}
\end{equation}

where $\sigma_t$ and $\sigma_r$ are the catalogued uncertainties of the target and reference sources respectively, and $\sigma_\Delta$ is the final positional uncertainty of the source offset $\Delta$ after adding a floor value, $f$, in quadrature. The optimal floor value was established by a goodness-of-fit analysis, calculating the reduced chi-squared statistic for observed and residual offsets across several randomly selected scans using different error floor values ranging from $0''$ to $0.5''$. A floor value of $0.4''$ was found to yield a reduced chi-squared value closest to 1, indicating an optimal fit. This floor value of $0.4''$ was also validated through comparisons between RACS-Low1 and other surveys, such as FIRST and VLASS, which showed similar results. This approach ensured that any non-zero systematic contributions to the positional uncertainties were accurately captured, preventing spurious corrections that might arise from underestimating the uncertainties in the individual source positions.

These crossmatching and modelling steps were reiterated for the residual offsets, now reducing the crossmatching threshold to $2''$. This refinement was aimed at eliminating any false positive crossmatches, further improving the precision of our modelling. By applying this more stringent threshold, we ensured that the remaining offsets reflected only genuine positional discrepancies, leading to a more accurate astrometric correction across the entire RACS sky coverage.

With approximately 600 million WISE sources in the sky compared to around 3 million RACS sources, only  $\sim 0.5\%$ of WISE sources have a RACS counterpart. Assuming a uniform source distribution across the entire sky, the probability of a false crossmatch for any RACS source with a $5''$ matching radius is $\sim 3\%$. This implies that for every 100 crossmatched sources, approximately 3 are likely spurious. By further refining the crossmatching radius to $2''$, the rate of spurious matches drops to under 0.4\%, an insignificantly low number that does not meaningfully impact our corrections. However, in regions on the Galactic plane, where source density is notably higher, the probability of false matches increases substantially, which likely contributes to the reduced reliability of our corrections in those areas.

The original RACS-Low1 catalogue was obtained using the methods described in \citealp{2021PASA...38...58H}. We generated a new catalogue with the corrected positional values by updating the RA and Dec. of each source, using positional corrections of up to three of the nearest beams within a 1-degree radius. The corrections were applied based on a weighted system, where closer corrections were given higher weight (as shown in Equation~\ref{eq:weightedavg}) in determining the final position of the source. This weighting system ensured that positional shifts were as precise as possible, minimising the impact of large, spurious offsets from more distant corrections. This method was employed to produce both the corrected source catalogue and the Gaussian component catalogue. 

\begin{equation} \label{eq:weightedavg}
    \overline{\Delta_{final}} = \frac{\sum_{i=1}^n \frac{1}{\theta_i} \Delta_i}{\sum_{i=1}^n \frac{1}{\theta_i}},  \quad \text{where } n \leq 3
\end{equation}

where, $\overline{\Delta_{final}}$ is the final weighted-average offset correction for a given source in the catalogue, $\Delta_i$ are the beam offset corrections nearest to the source, and $\theta_i$ is the angular separation of the $i$-th beam correction from the source. 



The final residual offsets were compared with FIRST, VLASS and RFC as a verification check. Each of these surveys was chosen for a different reason, as previously mentioned. The results for the entire survey are consolidated in Section~\ref{sec:racslow1corrections}.

\begin{figure*}[ht]
    \centering
    \includegraphics[scale=0.52]{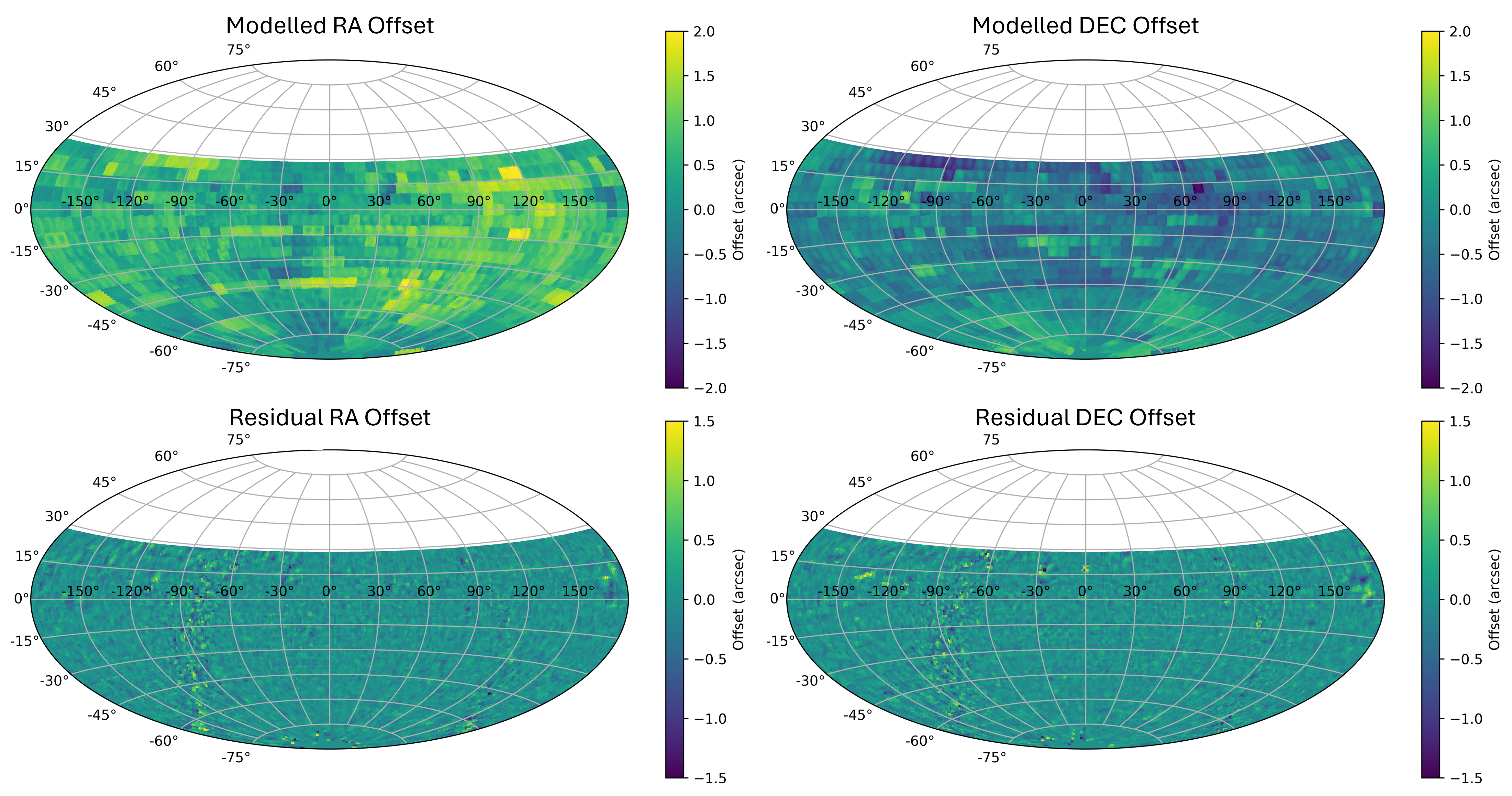}
    \caption{The modelled and residual offsets of RACS-Low1 vs WISE are shown here. The top row shows the modelled offsets and the bottom row shows the residual offsets in RA (left) and Dec. (right) for the entire sky coverage.}
    \label{fig:RACS-Low1aitoff}
\end{figure*}

\subsection{Modelling Parameters for RACS-Low3} \label{sec:modelRACS-Low3}
The methodology discussed in the previous Section~\ref{sec:modelracslow1} is further modified, improved and then extended to the new RACS-Low3 scans as well as the other RACS bands, which are detailed below.

We used the post-processed, publicly available version of the RACS-Low1 catalogue, where the scan tiles were convolved to a common resolution of $25''$ and mosaicked with neighbouring tiles. While this provided uniform resolution across the sky coverage, it introduced complications for astrometric modelling. Specifically, beams located at the edges of these tiles contained mosaicked source information from adjacent tiles, leading to positional uncertainties. These edge effects caused the astrometric offsets derived from RACS-Low1 to be less reliable in regions where the mosaicking blended sources from multiple tiles, diminishing the overall efficiency of the modelling process.


In contrast, the RACS-Low3 observatory project provided us with un-mosaicked source lists for each individual beam for each scan across the entire sky coverage. This was an improvement, as we were able to directly model the true positional offsets for each beam in each scan without contamination from neighbouring scans. The un-mosaicked data ensured that the offsets derived were accurate representations of each beam's intrinsic position, unpolluted by artefacts arising from neighbouring observations. This, combined with advancements in ASKAP instrumentation and refined observation techniques between the time of RACS-Low1 and RACS-Low3, we expected would result in better overall astrometry in RACS-Low3.

RACS-Low3 consists of 1,493 unique scans in the lowest frequency band (RACS-Low). It differs from RACS-Low1 in several key aspects. Firstly, it employs a more compact beam footprint, known as $closepack36$, as opposed to the $square\_6x6$ configuration used in RACS-Low1 (as detailed in Section 9.10 of \citealp{2021PASA...38....9H}). This updated footprint aligns RACS-Low3 observations with those of RACS-Mid1 and RACS-High1, enabling improved consistency across RACS epochs for different ASKAP bands. Additionally, RACS-Low3 uses a slightly higher observing frequency of 943.5 MHz compared to the 877.5 MHz used in RACS-Low1, where the ASKAP sensitivity is improved. In its raw state, each RACS-Low3 beam image is a square 4x4 degrees in size, capturing not only the main lobe of each PAF beam, but extending well out into the adjacent sidelobes. This is true for the pre-mosaicked beam images for all RACS epochs.


In our preliminary analysis of the RACS-Low3 data, we observed significantly larger and more erratic positional offsets in scans above Dec. $+30^\circ$ and within the Galactic plane. While the issues within the Galactic plane were anticipated due to reasons explained in Section~\ref{sec:modelracslow1}, for regions above Dec. $+30^\circ$, ASKAP's performance was likely degraded due to observations being conducted at very low elevations, where the the positional offsets exceeded $7''$ in some cases. These discrepancies can be attributed to three primary factors: the highly elongated point spread function (PSF) at low elevations, degraded observing conditions due to increased atmospheric distortion, and heightened solar activity impacting ionospheric stability. This resulted in missed crossmatches due to the limiting crossmatching threshold of our algorithm, which then affected the offset modelling. Better results at these declinations could be achieved by using a higher crossmatching radius. However, this approach would also introduce more false positives, reducing the reliability of the crossmatches. To address this, we plan to explore a declination-dependent crossmatching scheme as part of our future work (as discussed briefly in Section~\ref{sec:futureplans}).






Therefore, a crucial modification we implemented for the RACS-Low3 modelling process was to update the code to exclude both the Galactic plane and areas above Dec. $+30^\circ$ from the primary modelling process, that is, scans in this declination range did not contribute to the calculation of the scan-independent beam offsets. The exclusion of these problematic regions from the main modelling pipeline allowed us to focus on the parts of the sky where offsets followed more predictable patterns. This strategic omission improved the overall robustness and accuracy of the offset models, leading to more reliable corrections for the rest of the RACS-Low3 sky coverage. We adopted a different approach for scans in these regions, using the scan-independent beam offsets generated from the unaffected regions in the previous modelling steps and applying these to model the beam-independent scan offsets for the problematic areas. This approach allowed us to still model offsets in these regions without compromising the integrity of the broader dataset.

Since producing a mosaicked, uniformly corrected all-sky catalogue from the RACS-Low3 per-beam source lists using methods similar to those outlined in previous works such as \citet{2021PASA...38...58H} is beyond the immediate scope of our project, we tailored our workflow to meet the specific requirements of this study. We focused on combining the per-beam source lists, but with the updated RA and Dec. positions derived from the corrections generated for each source. This approach allowed us to refine the astrometry for the sources in RACS-Low3 in a way that directly addresses our scientific objectives. In a future work, we plan to make these corrected source lists available publicly, which can then be used to generate fully corrected RACS-Low3 catalogues.


\begin{figure*}[h!]
    \centering
    \includegraphics[scale=0.72]{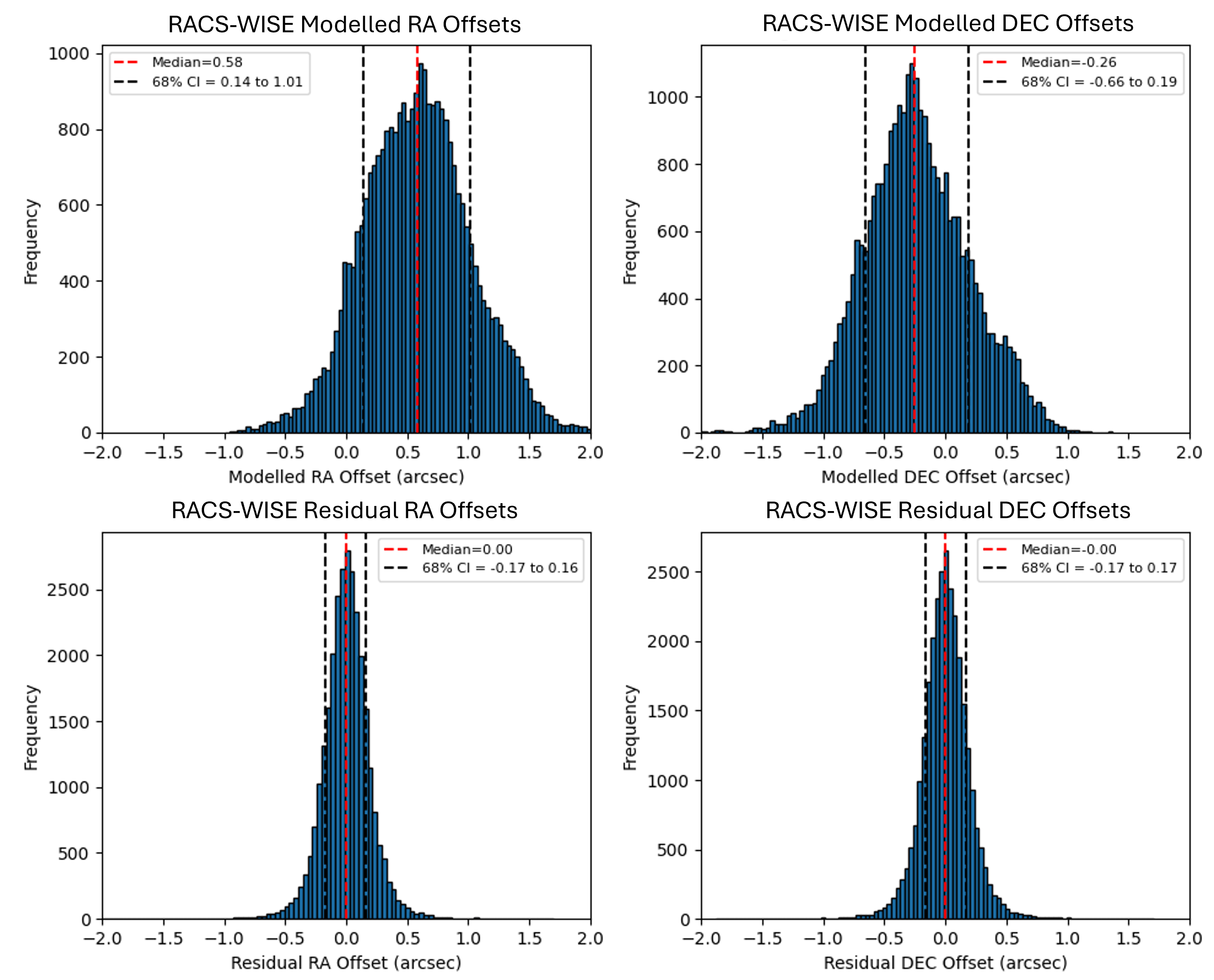}
    \caption{The modelled (top row) and residual (bottom row) offsets for RACS-Low1 vs WISE. The median RA offset (left) improves from $0.58''$ to $0.00''$ and the median Dec. offset (right) decreases from $0.26''$ to $0.00''$. The 68\% confidence intervals also narrow down significantly.}
    \label{fig:RACS-Low1histograms}
\end{figure*}

\section{Results} \label{sec:results}

\subsection{RACS-Low1 Corrections} \label{sec:racslow1corrections}
The observed and residual RACS--WISE offsets after corrections are shown in Aitoff projection plots in Figure \ref{fig:RACS-Low1aitoff} for $\sim$ 1.7 million crossmatched sources in $28980$ beams. The residual offsets in both RA and Dec. remain uniformly small across the entire sky coverage, with a few exceptions. These outliers are mainly located in the Galactic plane or in scans affected by very bright sources that contaminate the entire sky coverage or by observational data that is irreparably flawed (a few examples are discussed in Section~\ref{sec:excludedregions}). The median estimated uncertainty for both RA and Dec. is $0.12''$, and is calculated using a weighted mean of the documented positional errors from both RACS-Low1 and WISE. However, this method assumes that the positional correction is uniform across the entire sky coverage and that the RACS-Low1 and WISE source centroids align perfectly for every source, neither of which will be strictly true. As a result, while these values suggest that most RACS sources will have positional uncertainties below $0.3''$, they underestimate the true positional uncertainties. This is further discussed in Section~\ref{sec:racslow1rfc}.


To quantify these findings further, Figure \ref{fig:RACS-Low1histograms} provides histogram plots of the same parameters, with the median and 68\% confidence interval values indicated. As the histograms illustrate, the mean bias is completely eliminated in both RA and Dec. while the stochastic variation is also greatly reduced; the 68\% confidence intervals for the residual offset are reduced from $0.58''^{+0.43}_{-0.44}$ to $0.00''^{+0.16}_{-0.17}$ in RA and from $0.26''^{+0.45}_{-0.40}$ to $0.00''^{+0.17}_{-0.17}$ in Dec.



These results demonstrate that we have significantly improved the astrometry of RACS-Low1. To better quantify the residual errors and determine the appropriate uncertainty to assign to the corrected RACS catalogues, we now compare the corrected RACS positions with other suitable reference catalogues.

\begin{figure*}[h!]
    \centering
    \includegraphics[scale=0.72]{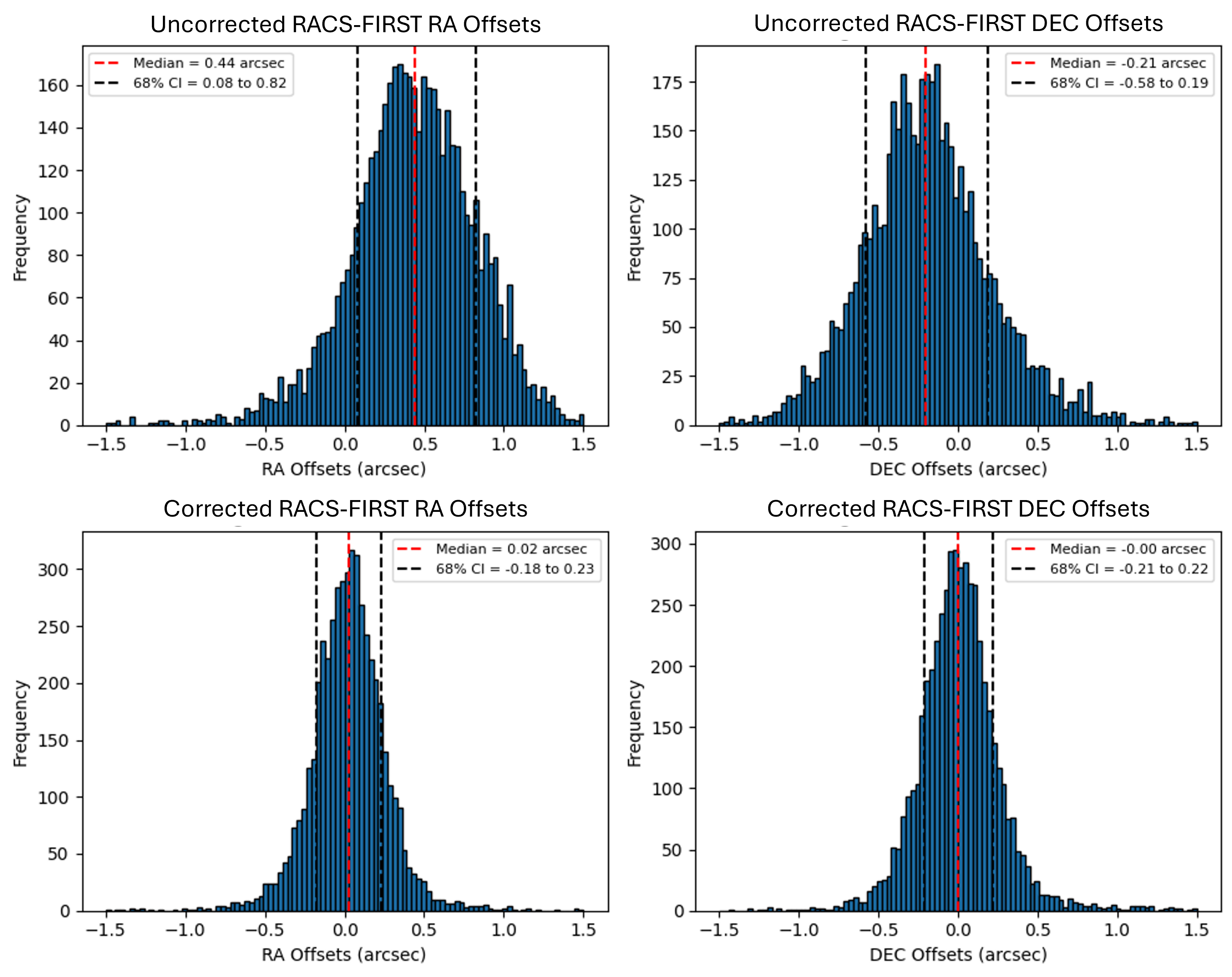}
    \caption{The uncorrected (top row) and corrected offsets (bottom row) in RA (left) and Dec. (right) for RACS-Low1 vs FIRST. The median offset values in RA and Dec. reduce from $0.44''$ and $-0.21''$ to $0.02''$ and $0.00''$ respectively.}
    \label{fig:RACS-Low1firsthist}
\end{figure*}

\begin{table*}[h!]
\centering
\caption{Astrometric Offset Comparison for Corrected RACS-Low1 with FIRST, VLASS, and RFC}
\begin{tabular}{|c|c|c|c|c|c|c|}
\hline
\textbf{Survey Comparison:} & \textbf{Total Sources} & \textbf{Median RA Offset} & \textbf{68\% Confidence Interval} & \textbf{Median Dec. Offset} & \textbf{68\% Confidence Interval} \\
\textbf{Corrected RACS-Low1 vs} & \textbf{Crossmatched} & \textbf{(arcseconds)} & \textbf{(RA, arcseconds)} & \textbf{(arcseconds)} & \textbf{(Dec., arcseconds)} \\ \hline

FIRST         &  $\sim250,000$  & 0.02   & -0.18 --- 0.23   & 0.00   & -0.21 --- 0.22  \\ \hline
VLASS (Total)         &  $\sim1,000,000$  & 0.04   & -0.18 --- 0.27   & -0.05  & -0.30 --- 0.21  \\ \hline
VLASS (Off-Plane)        &  $\sim900,000$  & 0.03   & -0.18 --- 0.26   & -0.05  & -0.29 --- 0.20  \\ \hline
VLASS (On-Plane)          &  $\sim100,000$  & 0.08   & -0.17 --- 0.37   & -0.06  & -0.36 --- 0.23  \\ \hline
RFC (Total)           &  $\sim12,000$  & 0.04   & -0.25 --- 0.34   & -0.03  & -0.34 --- 0.25  \\ \hline
RFC (Off-Plane)           &  $\sim11,000$  & 0.04   & -0.25 --- 0.33   & -0.03  & -0.33 --- 0.25  \\ \hline
RFC (On-Plane)            &  $\sim1,000$  & 0.03   & -0.29 --- 0.41   & -0.05  & -0.43 --- 0.26  \\ \hline

\end{tabular}
\label{tab:RACS-Low1othercat}
\end{table*}

\subsection{Verification of Corrections of RACS-Low1 using Other Catalogues} \label{sec:racslow1othercat}
We performed cross-verification of the RACS-Low1 astrometric corrections using comparisons with FIRST, VLASS, and RFC, and the results are tabulated in Table~\ref{tab:RACS-Low1othercat}.

\subsubsection{RACS-Low1 vs FIRST} \label{sec:racslow1first}

We selected the VLA-FIRST catalogue for verification checks despite its limited coverage of the RACS sky because it is the closest in both radio frequency and angular resolution of the available options. Therefore, while we cannot use FIRST to assess any spatial variations in the residual RACS-Low1 error, it is a the most suitable catalogue to evaluate the performance accurately in the regions it is available. By crossmatching RACS-Low1 sources with FIRST in the same beam-wise bins used for WISE, we generated histograms of the offsets in RA and Dec. Only $\sim 5000$ RACS-Low1 beams overlap with FIRST, with each beam having an average of 50 crossmatched sources. These histograms, shown in Figure~\ref{fig:RACS-Low1firsthist}, include offset distributions both before and after the corrections. Clear improvements can be seen from this figure and Table~\ref{tab:RACS-Low1othercat}, and the histograms also exhibit a more Gaussian distribution, which is the expected result of well-behaved astrometric corrections. These results will factor in to our estimate of the residual RACS-Low1 uncertainties in the off-plane region. Finally, we note that since the RACS--FIRST offsets are calculated for ensembles of sources across a small but finite patch of sky, these values may still slightly underestimate the true uncertainty for any individual RACS-Low source.




\begin{figure*}[h!]
    \centering
    \includegraphics[scale=0.72]{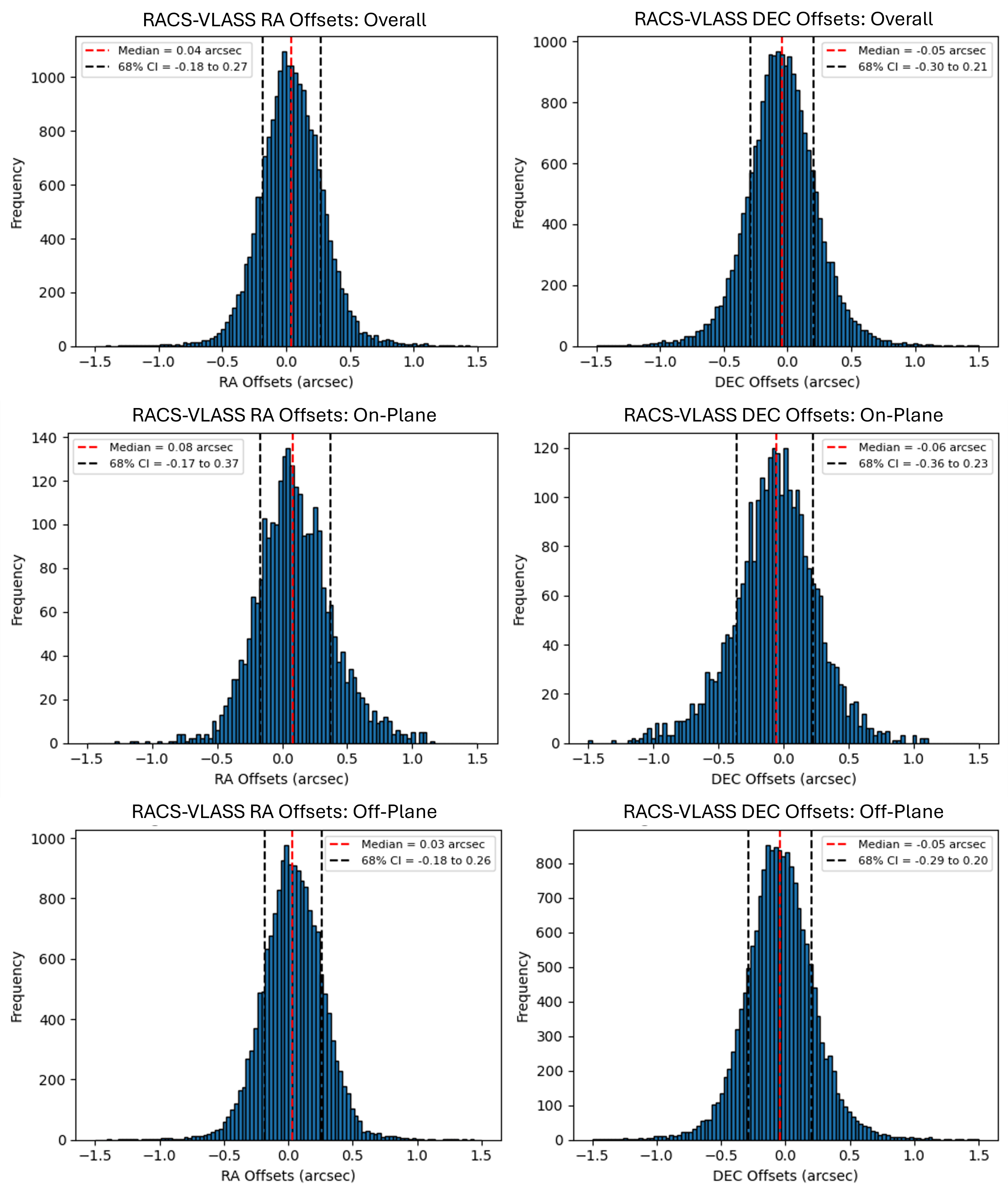}
    \caption{The corrected offsets in RA (left) and Dec. (right) for RACS-Low1 vs VLASS, with the overall histograms in the first row, the Galactic plane regions in the middle row, and the off-plane regions in the last row.}
    \label{fig:RACS-Low1vlasshist}
\end{figure*}

\subsubsection{RACS-Low1 vs VLASS} \label{sec:racslow1vlass}
Next, we compare RACS-Low1 with VLASS, that has a wider sky coverage than FIRST (down to a declination of $-40^\circ$). The higher observation frequency of VLASS will mean that for any sources that exhibit frequency-dependent structure, the centroid offsets from RACS-Low1 will be larger than was the case for FIRST. Moreover, the VLASS catalogue available on VizieR (\citealp{vizier2000}), which we used for verification, has known inherent all-sky offset of approximately $0.25''$ in declination (as documented in Section 3.3 of \citealp{2021ApJS..255...30G}), which we corrected prior to our crossmatching.  However, as can be seen in Figure 8 of the same article, there's a clear direction dependence on the Dec. offset, but it is unclear to what extent the spatial variations of VLASS residual errors have been mitigated by this single all-sky correction we employed (meaning that our single all-sky correction is likely to leave some residual spatially-dependent errors). $\sim 20000$ RACS-Low1 beams overlap with VLASS, with each beam having an average of 50 crossmatched sources. The median offsets and 68\% confidence intervals are shown in Figure~\ref{fig:RACS-Low1vlasshist} and Table~\ref{tab:RACS-Low1othercat}.





We further divide the dataset into on-plane and off-plane regions, to separate the more complex Galactic Plane region between Galactic Latitudes of $-10^\circ$ and $+10^\circ$, where we know the reliability of our model is lower. As can be seen in Table~\ref{tab:RACS-Low1othercat}, results are similar to FIRST in the off-plane regions. In the Galactic plane region, while the median RA and Dec. values are mostly similar, we see the histogram spread worsening in both RA and Dec., which shows that our model performs worse in these regions, as expected. As with the comparison to FIRST, the fact that ensemble source averages are used to derive these offsets means that the true residual RACS position uncertainty may be higher.


\begin{figure*}[h!]
    \centering
    \includegraphics[scale=0.72]{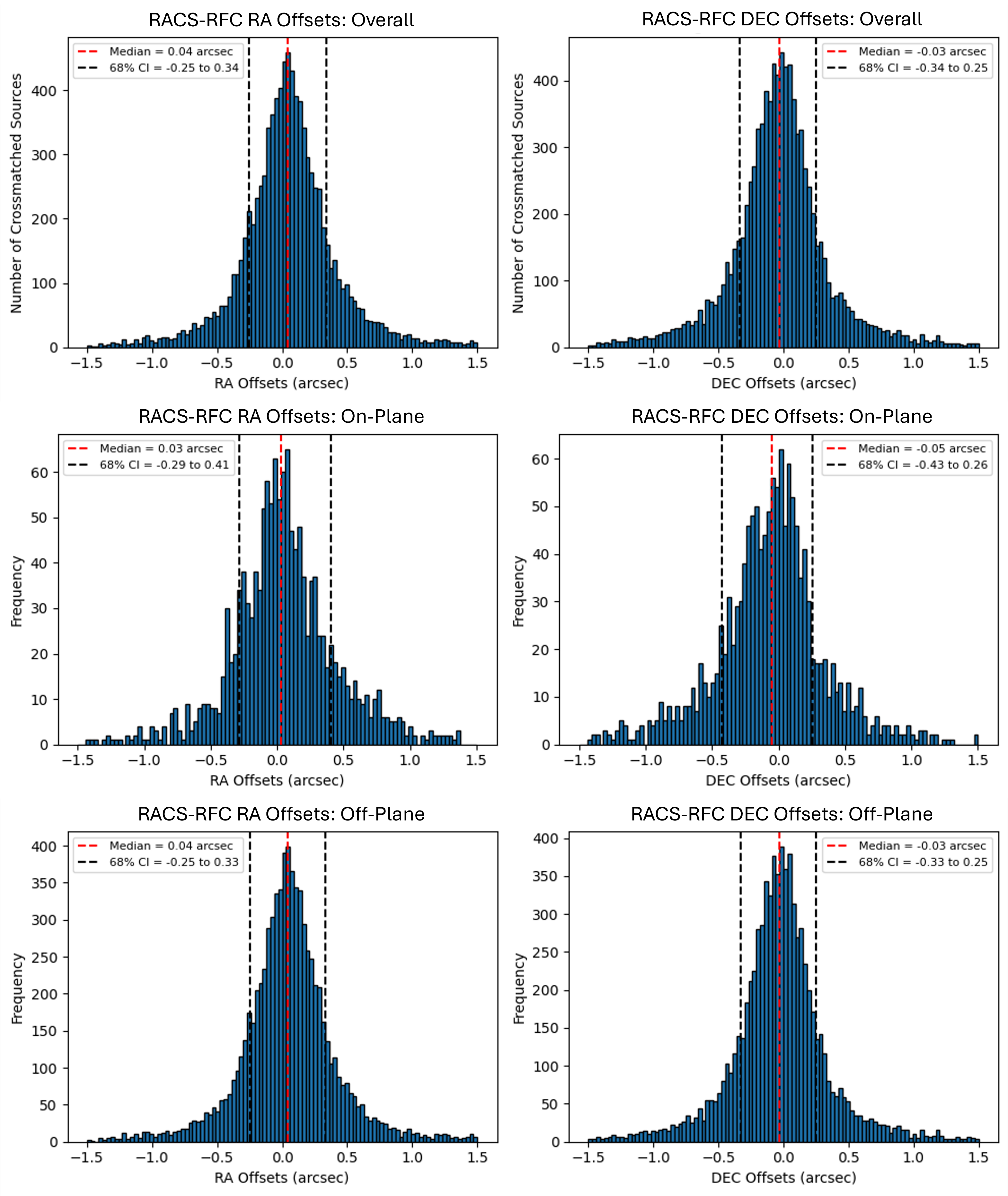}
    \caption{The corrected offsets in RA (left) and Dec. (right) for RACS-Low1 vs RFC, with the overall histograms in the first row, the Galactic plane region in the middle row, and the off-plane region in the last row. Of the $\sim 18,000$ RFC sources in the RACS-Low1 footprint, after the extended sources are filtered out, $\sim 12,000$ sources have a valid crossmatch to RACS-Low1 sources.}
    \label{fig:RACS-Low1rfchist}
\end{figure*}

\subsubsection{RACS-Low1 vs RFC} \label{sec:racslow1rfc}
Finally, we use RFC which is a VLBI all-sky catalogue with high astrometric precision. However, as it is restricted to relatively bright sources that are compact on mas scales, the source density is much lower than FIRST or VLASS. Due to this limitation, we deviate from the beam-wise approach used earlier and directly compare RACS-Low1 sources with RFC sources. Given this change in approach and RFC's higher frequency and resolution, the differences in centroid positions for extended sources would prevent per-source centroid offsets from averaging out, thereby negatively affecting our verification of RACS-Low1 corrections. As RFC does not have any information about the extent of the sources, we first crossmatch it with VLASS, in the region it is available, and filter out all the sources that correspond to extended sources in VLASS. 


We then crossmatch the remaining point-like RFC sources with the corrected RACS-Low1 catalogue, producing histograms as shown in Figure~\ref{fig:RACS-Low1rfchist}, with the median offsets and 68\% confidence intervals listed in Table~\ref{tab:RACS-Low1othercat}. To assess model performance in different regions, we again divided the dataset into on-plane and off-plane regions. In the Galactic plane region, the median RA and Dec. offsets are still within the margin of error, though the 68\% confidence intervals are noticeably broader. The histograms for the off-plane region align with the overall dataset, as expected, once again confirming the model's consistency outside the more complex Galactic plane region.



We further divided the dataset into different declination ranges to assess whether the performance of our astrometric model varied with declination. Across these ranges, the median RA offsets were found to vary between $0.00''$ and $0.10''$, while the median Dec. offsets ranged from $0.00''$ to $0.06''$. The 68\% confidence intervals for both RA and Dec. remained consistent, varying within $\pm0.40''$  from the median. However, we see that the model’s performance slightly deteriorates at Dec. above $+10^\circ$ and below Dec. $-60^\circ$, likely due to challenges related to lower elevation observations by ASKAP in these regions.

Since these values are measured using individual sources, any true centroid position differences contribute directly to the measured offset, meaning that these offsets should overestimate the true residual uncertainty in the corrected RACS positions (unlike the checks made using FIRST and VLASS). This provides a useful upper bound, complementing the beam-averaged approach with VLASS and FIRST. 


The comparison with FIRST provided 68\% confidence intervals of up to $\pm0.23''$ in RA and $\pm0.22''$ in Dec. The comparison with VLASS yielded 68\% confidence intervals of $\pm0.37''$ in RA and $\pm0.36''$ in Dec. for on-plane regions, while off-plane regions showed 68\% confidence intervals of $\pm0.26''$ in RA and $\pm0.29''$ in Dec. Similarly, the comparison with RFC gave us 68\% confidence intervals of up to $\pm0.41''$ in RA and $\pm0.43''$ in Dec. for on-plane regions, and $\pm0.33''$ for both RA and Dec. in off-plane regions. After carefully evaluating the underestimates from this methodology, and overestimates from single-source matches with RFC, we recommend assuming 1-$\sigma$ systematic uncertainties of $\pm0.30''$ for both RA and Dec. in regions outside the Galactic plane is suitable for RACS-Low1 corrections. For Galactic plane regions, where the model encounters more significant challenges due to complex source environments and fewer point sources, slightly higher 1-$\sigma$ systematic uncertainties of $\pm0.40''$ in both RA and Dec. are more appropriate.

The residual distributions are not completely Gaussian, with a (small) excess of fields at large errors (>3-$\sigma$). Specifically, $\sim 1.2\%$ of FIRST fields in RA and $\sim 2.1\%$ in Dec. (all of which are off-plane) have a residual error greater than $0.9'‘$ (3-$\sigma$), compared to the 0.3\% predicted for a Gaussian distribution. These deviations are dominated by a handful of fields in the {\em excluded regions} (as discussed in Section~\ref{sec:excludedregions}), where the RACS-Low1 data proved impossible to fully correct. For the vast bulk of the sky away from the Galactic plane, the post-correction uncertainty we prescribe should be appropriate: in the off-plane region, residual errors exceeding $0.6''$ (2-$\sigma$) are seen in just 3.0\% of FIRST fields for RA, and 4.3\% of FIRST fields for Dec. Comparable results are seen when comparing against VLASS in the off-plane region ($\sim 2.0\%$ fields in RA and $\sim 4.0\%$ in Dec. exceeded $0.6''$ residual offsets.) In the on-plane region, where we employ a more conservative estimate of 0.4’' for the 1-$\sigma$ uncertainty, no excess of VLASS field residuals is seen at either the 2-$\sigma$ ($\sim 2.3\%$ of fields in RA and $\sim 2.6\%$ of fields in Dec.) or 3-$\sigma$ ($\sim 0.2\%$ of fields in RA and $\sim 0.1\%$ in Dec.) level. Accordingly, while our recommended post-correction uncertainty for RACS-Low1 astrometry should be usable for almost the entire survey, we advise caution when using RACS-Low1 data from the {\em excluded regions}, all of which are detailed in Section~\ref{sec:excludedregions}.



\begin{figure*}[h!]
    \centering
    \includegraphics[scale=0.7]{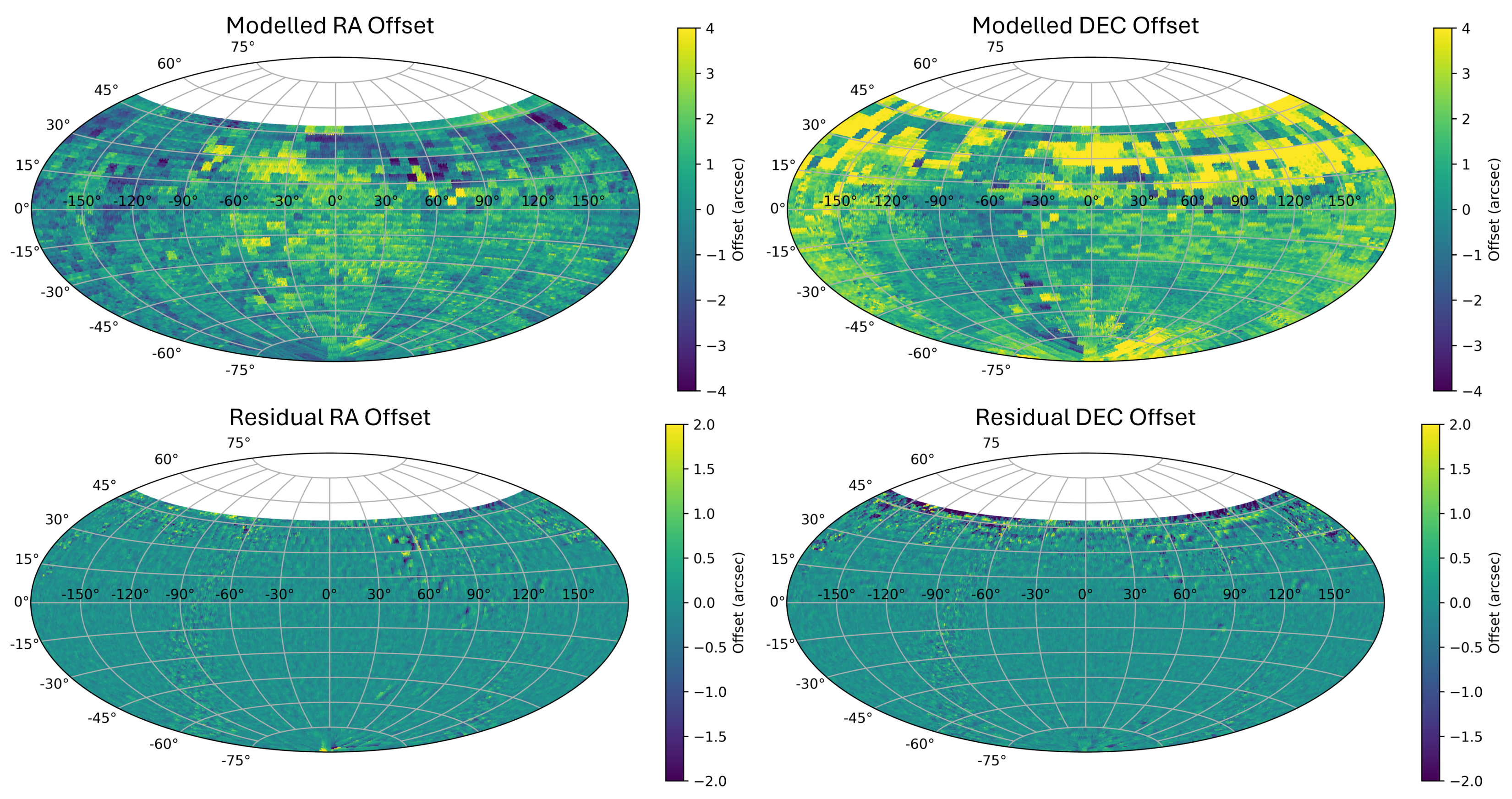}
    \caption{The modelled and residual offsets of RACS-Low3 vs WISE. The top row shows the observed offsets and the bottom row shows the residual offsets in RA (left) and Dec. (right) for the entire sky coverage.}
    \label{fig:RACS-Low3aitoff}
\end{figure*}

\begin{figure*}[h!]
    \centering
    \includegraphics[scale=0.72]{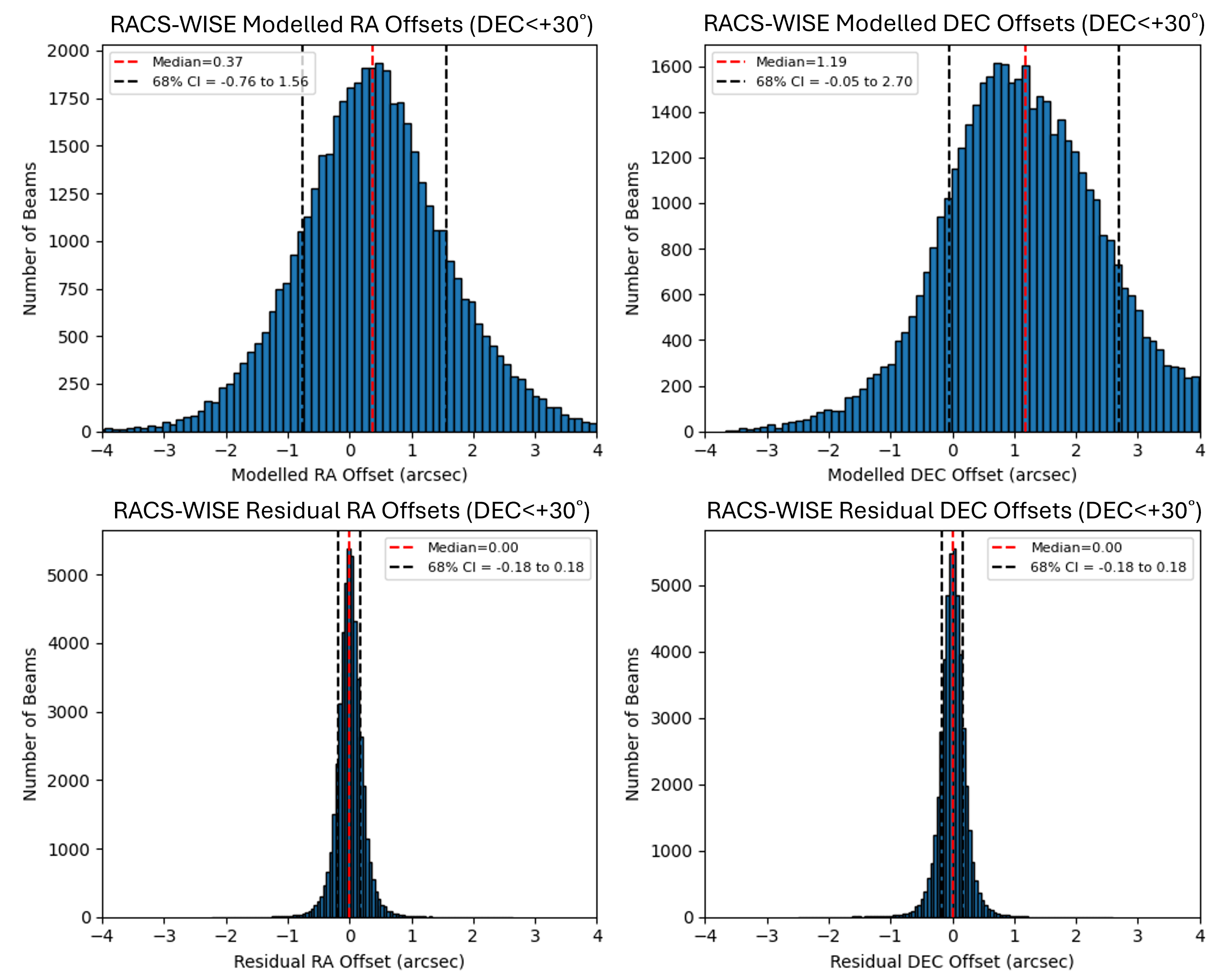}
    \caption{The modelled (top row) and residual (bottom row) offsets for RACS-Low3 vs WISE for scans below Dec. $+30^\circ$, to have a fair comparison between the corrections for RACS-Low3 and RACS-Low1. The median RA offset (left) improves from $0.37''$ to $0.00''$ and the median Dec. offset (left) decreases from $1.19''$ to $0.00''$. The 68\% confidence intervals also get significantly smaller.}
    \label{fig:RACS-Low3histograms}
\end{figure*}

\subsection{RACS-Low3 Corrections} \label{sec:racslow3corrections}
After completing the corrections for RACS-Low1, we applied an upgraded workflow to correct the 1493 un-mosaicked RACS-Low3 scans (53748 beams) using the same WISE data. The modelled and residual RACS-WISE offsets for RACS-Low3 are represented in Aitoff projection plots in Figure \ref{fig:RACS-Low3aitoff}. 

Several key points arise from preliminary inspection. First, the RACS-Low3 scans exhibit significantly denser coverage and are more uniformly distributed, extending over the entire sky below Dec. $+49^\circ$. However, the pre-correction offsets in both RA and Dec. are considerably larger than those in RACS-Low1 even in regions common to both these surveys. For regions above Dec. $+30^\circ$, which are unique to RACS-Low3, pre-correction offsets sometimes exceed $7''$. Post-correction, these offsets are markedly reduced across most of the sky, demonstrating the effectiveness of our model. The model’s performance is notably weaker in the Galactic plane, as observed with RACS-Low1, as well as in regions above Dec. $+30^\circ$, due to the inherent limitations of ASKAP at low elevations, such as elongated PSFs and atmospheric distortions.





We recognise that the performance of ASKAP degrades significantly above Dec. $+30^\circ$, so we first calculate the results for scans below Dec. $+30^\circ$, shown quantitatively in Figure~\ref{fig:RACS-Low3histograms}. As the histograms illustrate again, the mean bias is fully eliminated in both RA and Dec. while the stochastic variation is also significantly reduced; the 68\% confidence intervals for the residual offset are reduced from $0.37''^{+1.19}_{-1.13}$ to $0.00''^{+0.18}_{-0.18}$ in RA and from $1.19''^{+1.51}_{-1.24}$ to $0.00''^{+0.18}_{-0.18}$ in Dec. This shows that the astrometric accuracy as estimated by the WISE residuals, appears consistent between RACS-Low1 and RACS-Low3 in these regions. This implies that, somewhat surprisingly, the mosaicking in the RACS-Low1 catalogue did not have a significantly detrimental effect on our ability to correct the catalogue positions. 


When calculating the results for the entire RACS-Low3 sky, including the regions above Dec. $+30^\circ$, we similarly find that the mean bias is completely eliminated and the 
stochastic variation is also reduced; the 68\% confidence intervals for the residual offset are reduced from $0.29''^{+1.22}_{-1.19}$ to $0.00''^{+0.20}_{-0.20}$ in RA and from $1.24''^{+1.76}_{-1.29}$ to $0.00''^{+0.21}_{-0.21}$ in Dec. These results are noticeably worse because of the poor corrections in regions above Dec. $+30^\circ$, due to reasons discussed above.



\begin{figure}[h!]
    \centering
    \includegraphics[scale=0.65]{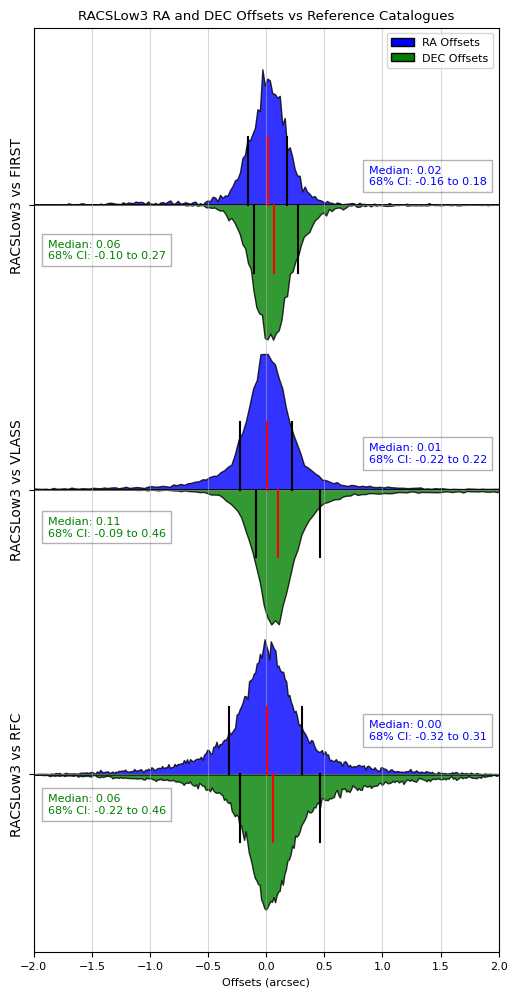}
    \caption{The corrected offsets in RA and Dec. for RACS-Low3 vs reference catalogues, with FIRST comparisons in the top row, VLASS in the middle row and RFC in the bottom row.}
    \label{fig:RACS-Low3firstvlassrfchist}
\end{figure}

\subsection{Verification of Corrections of RACS-Low3 using Other Catalogues} \label{RACS-Low3othercat}
We performed cross-verification of the RACS-Low3 astrometric corrections using comparisons with FIRST, VLASS, and RFC, collapsing the data into a single analysis for simplicity, as shown in Figure~\ref{fig:RACS-Low3firstvlassrfchist}. Crossmatching RACS-Low3 with FIRST resulted in median offsets in RA and Dec. improving from $0.19''$ and $1.13''$ (uncorrected) to $0.02''$ and $0.06''$ (corrected). Owing to the limited coverage of FIRST, only $\sim 9000$ RACS-Low3 beams are included, with each beam having an average of 50 crossmatched sources.

For the comparison of RACS-Low3 with VLASS, we used the latest catalogues available on CIRADA\footnote{Website: \url{cirada.ca/vlasscatalogueql0}}, the official repository for VLASS data. These updated catalogues include corrections for the inherent Dec. offset observed in previous versions of VLASS. This produced median RA and Dec. offsets of $0.01''$ and $0.11''$, respectively. $\sim 42000$ beams of RACS-Low3 have overlapping sky coverage with VLASS, with each beam having an average of 50 crossmatched sources. There were persistent outliers in Dec. offsets, primarily within the Galactic plane and Dec. above $+20^\circ$, which contributed to larger Dec. offsets and indicate a continued challenge for accurate astrometric correction in these regions. 

Crossmatching with the high-precision RFC catalogue yielded median RA and Dec. offsets of $0.00''$ and $0.06''$ for $\sim 14000$ crossmatched sources. Although the RA offsets met expectations, the Dec. offsets exhibited a larger median than anticipated, mostly because of the poor corrections in regions above a Dec. of $+20^\circ$. These results suggest that while the corrections improved the overall astrometric accuracy of RACS-Low3, further refinement is necessary, particularly in challenging regions like the Galactic plane and higher declinations.

\begin{figure}[h!]
    \centering
    \includegraphics[scale=0.65]{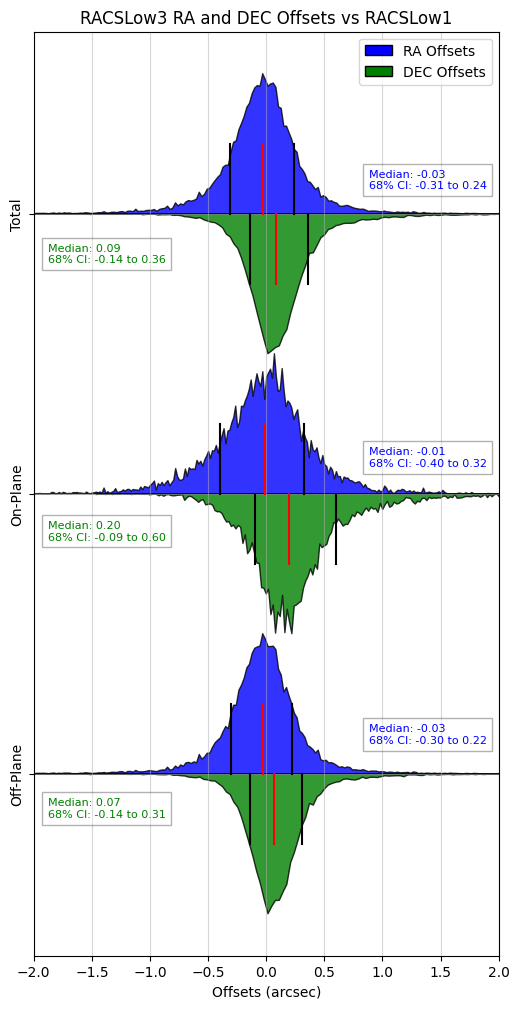}
    \caption{The offsets in RA and Dec. for RACS-Low3 vs RACS-Low1 post corrections, with the overall histograms in the top row, the on-plane region in the middle row, and the off-plane region in the bottom row. The overall histograms only include regions below Dec. $+30^\circ$, to compare common data available for these surveys, which corresponds to $\sim 46000$ RACS-Low3 beams, with each beam having an average of 50 crossmatched sources.}
    \label{fig:RACS-Low3RACS-Low1hist}
\end{figure}

\subsection{RACS-Low3 vs RACS-Low1}
\begin{figure*}[ht]
    \centering
    \includegraphics[scale=0.52]{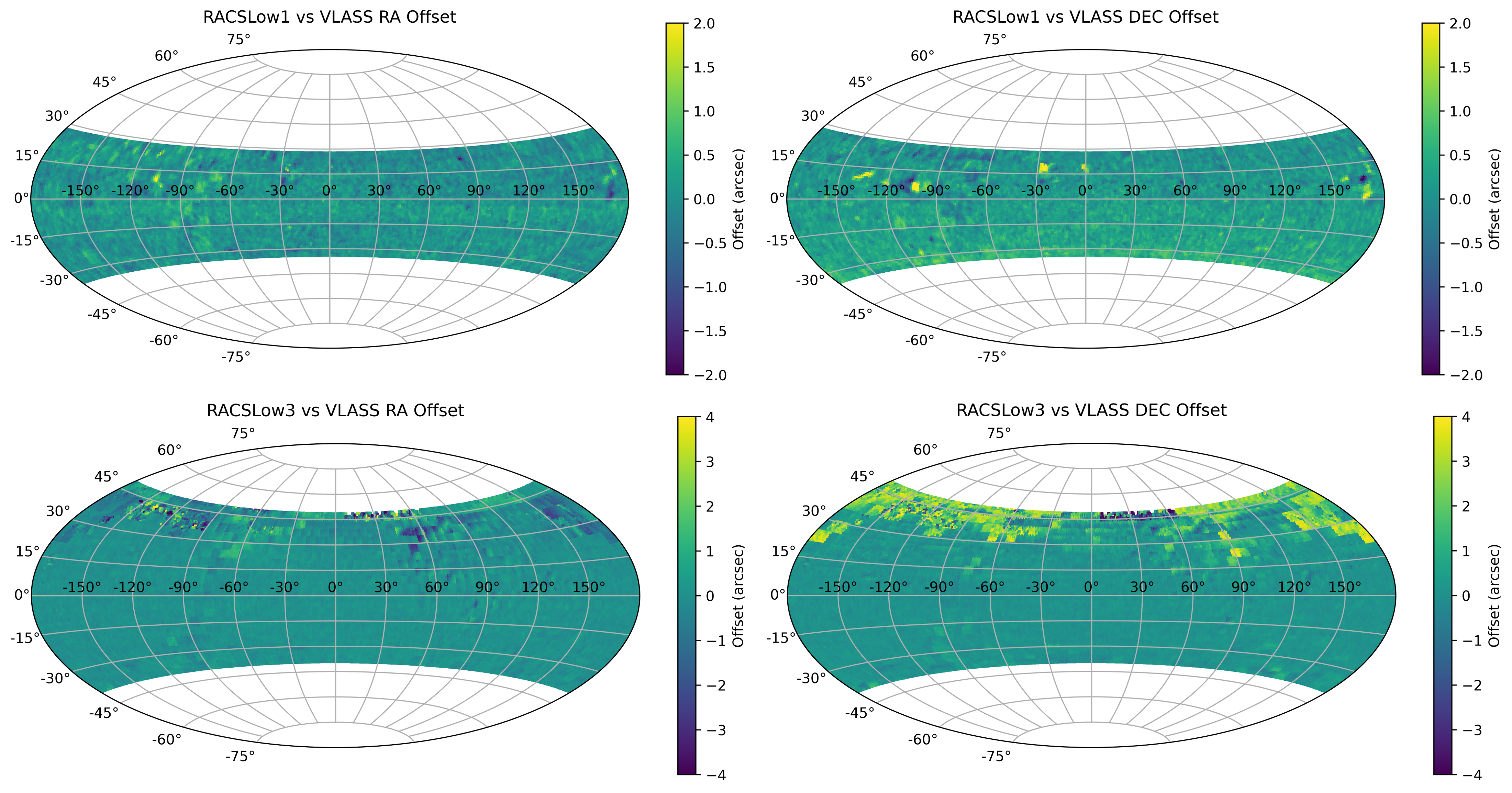}
    \caption{The top row shows the RA (left) and Dec. (right) offsets of RACS-Low1 vs VLASS and the bottom row shows the RA (left) and Dec. (right) offsets of RACS-Low3 vs VLASS. The underperforming regions in both RACS-Low1 and RACS-Low3 can be clearly visualised from these plots.}
    \label{fig:RACS-Low13vlassaitoff}
\end{figure*}

As a final sanity check, we crossmatched the corrected sources in RACS-Low3 with those in RACS-Low1. The histograms in Figure~\ref{fig:RACS-Low3RACS-Low1hist} show the median offsets and their 68\% confidence intervals of $-0.03''^{+0.27}_{-0.28}$ in RA and $0.09''^{+0.27}_{-0.23}$ in Dec. This comparison demonstrates that the corrected sources in RACS-Low3 are, within the margin of error, in the same positions as the corrected sources in RACS-Low1. However, in a few outlier regions (as evidenced in the residual plots shown in Figure \ref{fig:RACS-Low13vlassaitoff}), our model does not perform optimally due to the significantly lower number of point sources in each beam. 



Following these cross-verifications, while our model improves the positional accuracy of RACS-Low3, the larger number of beams that are less effectively corrected—-particularly in regions within the Galactic Plane and at northern declinations above Dec. $+30^\circ$—-indicates that the overall reliability of the final RACS-Low3 corrections is lower compared to RACS-Low1. This highlights the need for further refinement of the model, especially in areas with fewer point sources, to ensure optimal accuracy across the entire RACS-Low3 sky coverage. Given these results, we propose not to use RACS-Low3 for further processing at this stage and to continue exclusively with RACS-Low1.

\section{Discussion} \label{sec:discussion}


The encouraging outcome of our present focused exploration to correct the positional offsets in RACS-Low, paves way to proceed to the next step to generate a catalogue of corrected positions of all continuum sources across the entire ASKAP frequency band. Learning from the challenges we faced during our present endeavour, we aim to develop a more efficient and adaptive pipeline to model the offsets in RACS-Mid and RACS-High, as well as any future ASKAP survey.

\subsection{Accuracy of Corrections and Excluded Regions} \label{sec:excludedregions}

A quick analysis of the residual offsets of RACS-Low1, as depicted in Figure~\ref{fig:RACS-Low1aitoff}, highlights several regions where the astrometric corrections are less reliable than expected. Notably, the Galactic plane exhibits poor corrections, which is further elaborated in Section~\ref{sec:modelracslow1}. Additionally, certain {\em excluded regions} present significantly flawed corrections, as also seen in the RACS-Low1--VLASS offsets in the top row of Figure~\ref{fig:RACS-Low13vlassaitoff}.

For instance, Hercules A, a very bright radio source located at RA $16^{h}51^{m}$ and Dec. $+04^{\circ}59'$, corrupted the original data, likely due to its sidelobes, which shift the centroids of real sources by layering noise over the top, leading to unreliable corrections within a $3^\circ$ radius. Similarly, the Crab Nebula at RA $05^{h}34^{m}$ and Dec. $+22^{\circ}01'$ also caused contamination, resulting in poor corrections within a similar radius. Additionally, regions at RA $22^{h}12^{m}$ and Dec. $+16^{\circ}50'$, at RA $11^{h}20^{m}$ and Dec. $+05^{\circ}00'$, and at RA $00^{h}00^{m}$ and Dec. $+17^{\circ}00'$, show significant astrometric errors despite the absence of known sources or external factors. In these areas, corrections are highly unreliable, and uncertainties are estimated at $\pm0.50''$ for RA and $\pm1.00''$ for Dec.

Given the magnitude of these uncertainties, we suggest that for radio crossmatching/astrometric purposes, it would be preferable to not use RACS-Low1 in these regions. Instead, alternate catalogues like VLASS or RACS-Low3, which shows much more reliable corrections (as evidenced by the residual plots in Figure~\ref{fig:RACS-Low3aitoff} and RACS-Low3--VLASS offsets in the bottom row of Figure~\ref{fig:RACS-Low13vlassaitoff}), should be considered.




\begin{figure*}[h!]
    \centering
    \includegraphics[scale=0.31]{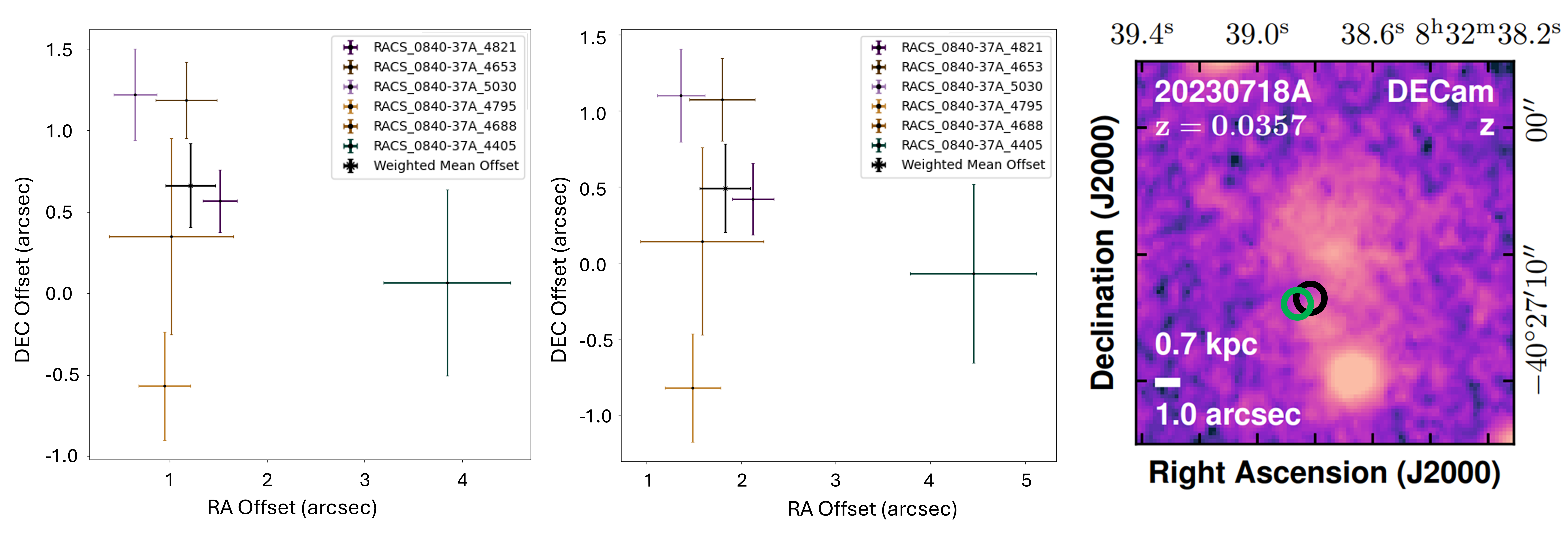}
    \caption{An example of determining the positional offsets of FRB20230718A while using the uncorrected (left) and the corrected (middle) RACS-Low1 catalogues with CELEBI. After updating the RACS catalogue, the estimated offset correction based on the field sources changed from $1.21'' \pm 0.26''$, $0.66'' \pm 0.26''$ to $1.75'' \pm 0.27''$, $0.49'' \pm 0.29''$ in RA and Dec. respectively. After including the assumed 1-$\sigma$ systematic uncertainty of $0.4''$ (as the Galactic latitude of this source is $-0.37^\circ$) and the statistical FRB position uncertainty, the final position of the FRB is updated from $08^{h}32^{m}38^{s}.82 \pm 0.54''$ to $08^{h}32^{m}38^{s}.86 \pm 0.52''$ in RA and from $-40^{\circ}27'06.78'' \pm 0.56''$ to $-40^{\circ}27'06.95'' \pm 0.53''$ in Dec. after corrections. The effect of this change on the FRB's localisation within the host galaxy is shown in the DECam image of the host (right), where the black error ellipse indicates the localisation with the uncorrected RACS-Low1 catalogue, and the green ellipse shows the improved localisation achieved with the corrected version (\citealp{2024arXiv240802083S}).} \label{celebiex}
\end{figure*}

\subsection{Present use of the RACS-Low1 Corrections} \label{sec:presentuse}
The corrected RACS-Low1 catalogues are currently being utilised in the CRAFT Effortless Localisation and Enhanced Burst Inspection pipeline (CELEBI, \citealp{2023A&C....4400724S}) to refine the positions of ASKAP-discovered FRBs. A comprehensive review of this integration, along with other updates to the pipeline, will be discussed in \citealp{celebi2.0}. As an example, Figure~\ref{celebiex} shows the positional offsets of FRB20230718A (a detailed review of the methodology is presented in \citealp{2024ApJ...962L..13G}) before and after using the corrected RACS-Low1 catalogue in CELEBI. The change in the source position can be visualised from the DECam\footnote{Website: \url{noirlab.edu/public/programs/ctio/victor-blanco-4m-telescope/decam/}} image of the FRB field. Additionally, a list of FRBs discovered using the ICS system, with their positions updated using the corrected RACS-Low1 catalogue, is presented in \citealp{2024arXiv240802083S}.

\subsection{Future Plans} \label{sec:futureplans}
Building on the astrometric improvements with RACS-Low1 and RACS-Low3, we plan to extend our corrections to the full RACS catalogue, including RACS-Mid and RACS-High. By applying similar crossmatching techniques with catalogues like WISE, FIRST, VLASS, as well as the corrected RACS-Low1, we aim to improve astrometric accuracy across all ASKAP frequencies. As with RACS-Low3, we will use the un-mosaicked per-beam source lists for both RACS-Mid and RACS-High to avoid the blending of astrometric errors that occur in the tile imaging process. Learning from our experiences in correcting RACS-Low1 and RACS-Low3, we will optimise our correction algorithm for future iterations. By first crossmatching with the corrected RACS-Low catalogues, we can apply larger crossmatching radii in regions with large positional offsets without significantly increasing the risk of false-positive matches, by avoiding the need to use excessively large radii directly with the dense WISE catalogue, where source confusion is more likely. This will help us generate enhanced corrections not only in the Galactic plane but also for areas at higher declinations. 

We also plan to investigate the impact of ionospheric stability on positional offsets in the different RACS epochs, particularly for observations conducted at low elevations. Given the heightened solar activity during some of these observations, ionospheric fluctuations may have introduced additional astrometric distortions. A potential direction for future work is to analyse time-dependent trends, such as diurnal or seasonal variations in positional errors, to assess whether ionospheric conditions contribute systematically to the observed discrepancies. This could inform more refined corrections, especially for fields observed under varying ionospheric conditions.



This unified, corrected RACS catalogue will serve as a reliable foundation for precise localisation for declinations below $+20^\circ$, the first of its kind at this resolution in the Southern Hemisphere, offering over 5 times the astrometric accuracy compared to the previous metric (SUMSS). It will complement VLASS, which already provides a reliable localisation catalogue in the Northern Hemisphere. Together, these resources will enhance future radio observations, particularly for FRB studies, by offering comprehensive astrometric accuracy across both hemispheres.

\section{Conclusion} \label{sec:conclusion}
In this paper, we developed and applied a robust astrometric correction model for the RACS-Low1 and RACS-Low3 radio continuum surveys conducted with ASKAP. By crossmatching RACS sources with the WISE catalogue, we were able to produce accurate positional corrections for both RACS-Low1 and RACS-Low3, significantly improving their astrometric precision. For RACS-Low1 corrections with WISE, our model demonstrated substantial improvements, with median RA and Dec. offsets reducing to within $0.00''$ and the 68\% confidence intervals were calculated to within $\pm0.30''$ all across the RACS-Low1 sky coverage, except the Galactic plane, where they were slightly worse at $\pm0.40''$. This consolidation of the corrected offsets allowed for a significant enhancement in the astrometric accuracy of the RACS catalogue, which in turn improved the precision of FRB localisations.

We validated our correction model through cross-comparisons with other high-precision catalogues such as FIRST, VLASS, and RFC. These comparisons confirmed the effectiveness of our model, with residual offsets for RACS-Low1 well within acceptable limits. However, the model showed some limitations in regions with sparse point sources, such as the Galactic plane, where the model performed less effectively.

RACS-Low3, being an unprocessed catalogue at the time of this study, posed additional challenges, but we successfully implemented our correction model using a similar methodology. While RACS-Low3 initially presented larger offsets than RACS-Low1, especially above Dec. +30°, the corrected positions showed marked improvements across most of the sky. However, due to the larger number of beams in RACS-Low3 that were corrected less optimally, the final astrometric accuracy was less reliable than that of RACS-Low1. Cross-verifications between RACS-Low1 and RACS-Low3 further demonstrated the reliability of the corrected positions, although certain outlier regions remained where the model did not perform as well, largely due to the limited number of point sources available for crossmatching or inherent limitations of the ASKAP telescope.


We also plan to release the corrected RACS-Low1 catalogues and RACS-Low3 per-beam source lists publicly, along with all the corrections applied, making them available for use by the broader scientific community. These corrections will not only mitigate the inherent astrometric uncertainties in RACS-Low epochs but also provide a framework for future improvements in radio surveys, ensuring that ASKAP continues to deliver cutting-edge data for the study of radio transients, and a wide range of other astronomical phenomena. This work paves the way for future astrometric corrections in other ASKAP surveys, including RACS-Mid and RACS-High, where we plan to develop a more adaptive pipeline that addresses the challenges encountered in this study.

By leveraging these meticulous corrections, we enable robust associations between FRBs and sources in optical catalogues, facilitating the identification of their multiwavelength counterparts. This precision is critical for pinpointing host galaxies, determining their redshifts, and uncovering details about their environments. Beyond FRBs, improved astrometry supports the study of other transient phenomena. For instance, accurate localisation aids in identifying optical counterparts of flare stars, enabling spectral type classification and further characterisation of their physical properties. Similarly, for LPTs, better positional precision can reveal potential optical companions or help constrain their nature.

\begin{acknowledgement}
The authors thank Simon C.-C. Ho for comments on the manuscript. AJ, ATD, and YW acknowledge the support of the Australian Research Council (ARC) grant DP220102305. MG is supported by the Australian Government through the Australian Research Council’s Discovery Projects funding scheme (DP210102103), and through UK STFC Grant ST/Y001117/1. For the purpose of open access, the author has applied a Creative Commons Attribution (CC BY) licence to any Author Accepted Manuscript version arising from this submission. We acknowledge the use of OpenAI's ChatGPT-4o (https://chatgpt.com/) for providing suggestions on refining sentence structure in this manuscript. This scientific work uses data obtained from Inyarrimanha Ilgari Bundara / the Murchison Radio-astronomy Observatory. We acknowledge the Wajarri Yamaji People as the Traditional Owners and native title holders of the Observatory site. CSIRO’s ASKAP radio telescope is part of the Australia Telescope National Facility (https://ror.org/05qajvd42). Operation of ASKAP is funded by the Australian Government with support from the National Collaborative Research Infrastructure Strategy. ASKAP uses the resources of the Pawsey Supercomputing Research Centre. Establishment of ASKAP, Inyarrimanha Ilgari Bundara, the CSIRO Murchison Radio-astronomy Observatory and the Pawsey Supercomputing Research Centre are initiatives of the Australian Government, with support from the Government of Western Australia and the Science and Industry Endowment Fund. This paper includes archived data obtained through the CSIRO ASKAP Science Data Archive, CASDA (https://data.csiro.au). A part of this work was performed on the OzSTAR national facility at Swinburne University of Technology. The OzSTAR programme receives funding in part from the Astronomy National Collaborative Research Infrastructure Strategy (NCRIS) allocation provided by the Australian Government. This research has made use of the VizieR catalogue access tool, CDS, Strasbourg, France \citealp{10.26093/cds/vizier}. The original description of the VizieR service was published in \citealp{vizier2000}. The Canadian Initiative for Radio Astronomy Data Analysis (CIRADA) is funded by a grant from the Canada Foundation for Innovation 2017 Innovation Fund (Project 35999) and by the Provinces of Ontario, British Columbia, Alberta, Manitoba and Quebec, in collaboration with the National Research Council of Canada, the US National Radio Astronomy Observatory and Australia’s Commonwealth Scientific and Industrial Research Organisation.
\end{acknowledgement}

\paragraph{Data Availability Statement}
The updated RACS-Low1 sources-only and Gaussian components catalogues, which include the astrometric corrections applied in this work, will be made available in CSV format through the PASA datastore (https://dmc.datacentral.org.au/institute/pasa). Researchers and interested parties will be able to access these catalogues to facilitate further studies. Details on accessing the dataset will be included in the online supplementary materials of this publication.


\printendnotes

\printbibliography




\end{document}